\documentclass[conference]{IEEEtran}
 \makeatletter
 \def\ps@headings{%
 \def\@oddhead{\mbox{}\scriptsize\rightmark \hfil \thepage}%
 \def\@evenhead{\scriptsize\thepage \hfil \leftmark\mbox{}}%
 \def\@oddfoot{}%
 \def\@evenfoot{}}
 \makeatother

\usepackage{graphicx}
\usepackage{amssymb}

\usepackage{algorithm}
\usepackage{algorithmic}

\newcommand{\mynote}[1]{{{\medskip
\footnotesize \em \noindent Note: #1}}\medskip}

\newcommand{\mynotex}[1]{}

\renewcommand{\mynote}[1]{}

\newcommand{\M}{\mbox{\tiny M}}
\newcommand{\Leaf}{\mbox{\tiny L}}
\newcommand{\R}{\mbox{\tiny R}}
\newcommand{\wifi}{\mbox{\tiny wifi}}
\newcommand{\bkhl}{\mbox{\tiny bkhl}}
\newcommand{\mmid}{\mbox{\tiny mid}}
\newcommand{\MC}{\mbox{\emph{MC}}}
\newcommand{\TC}{\mbox{\emph{TC}}}

\begin{document}

\title{Efficient Proactive Caching for Supporting \\ Seamless Mobility
\vspace{-0.4in}
}

\author{
\IEEEauthorblockN{Vasilios A. Siris, Xenofon Vasilakos, and George C. Polyzos}
\IEEEauthorblockA{
Mobile Multimedia Laboratory, Department of Informatics \\
Athens University of Economics and Business, Greece\\
\{vsiris, xvas, polyzos\}@aueb.gr
}
}

\maketitle

\begin{abstract}
We present a distributed proactive caching approach that exploits user mobility information to decide where to proactively cache data to support seamless mobility, while efficiently utilizing cache storage using a congestion pricing scheme. The proposed approach is applicable to the case where objects have different sizes and to a two-level cache hierarchy, for both of which the proactive caching problem is hard. Additionally, our modeling framework considers the case where the delay is independent of the requested data object size and the case where the delay is a function of the object size. Our evaluation results show how various system parameters influence the delay gains of the proposed approach, which achieves robust and good performance relative to an oracle and an optimal scheme for  a flat cache structure.

\mynotex{Possible also include that we prove they are NP?}

\end{abstract}

\vspace{-0.09in}
\section{Introduction}
\vspace{-0.03in}

\mynote{Motivation:
\begin{itemize}
\item Seamless mobility is important. Proactive caching or prefetching can be used to reduce the time for a mobile user to obtain his requested data.

\item Information for predicting a user/devices's  mobility is available.

\item Information of data a mobile/user requests is available, e.g. in ICN, or can be predicted (e.g. based on previous requests).

\item Two-level cache hierarchy: caches at the lower level can be located in e.g., hotspots, that belong to a particular provider or organization. The second level cache can be located close to the link interconnecting the provider or organization's network to the Internet.

\item Why not more than two-levels?

\item The concept of proactive caching is not new.
Problem we address is to efficiently utilize storage of caches that proactively fetch data, while achieving high performance in terms of reducing average delay. Also, approach is applied for variable data object sizes and a two-level hierarchy, and considers case where delay is independent and case where it depends on the object size.
\end{itemize}

}

Proactively fetching data  to reduce delay in mobility scenarios  has been proposed within the context of publish-subscribe network architectures  \cite{burcea2004disconnected,gaddah2010extending,siris2011selective}, for vehicular WiFi access \cite{Des++09}, and more recently for cellular networks \cite{Gol++12,Mal++12}. Even earlier, prefetching has been proposed and investigated for file-system and Web access performance enhancement at least since the mid nineties. In this paper we focus on proactive caching related to mobility.

Proactive caching  achieves gains by making data objects requested by a mobile user immediately available when the mobile moves to a new network attachment point, thus reducing the delay for obtaining the data compared to transferring it from the original source where the data is located. The higher delay for obtaining the data from the source can be due to the larger distance (number of hops), or because the path to the source includes low capacity links, such as a low capacity backhaul  of femto/small cells or WiFi hotspots.
Proactive caching can exploit mobility information, such as the probability of a  mobile connecting to  future  attachment points. Exploiting such mobility information to undertake proactive actions has been applied to reduce handover delays  in WiFi and cellular networks \cite{chiu2000predictive,soh2003qos,P++05}.

We propose a novel distributed approach for proactively fetching  data that a mobile requests in caches located close the network attachment points where the mobile has some probability to connect to.
The caches that should proactively fetch data objects are selected based on mobility information and on the gains from reduced data transfer delay.
The approach is applicable to the case where the requested data objects have different sizes and to a  two-level cache hierarchy, which are both hard problems. Moreover, our modeling framework considers the case where a data object's transfer delay is independent of the object's size and the case where the transfer delay is a function of the object's size.
Although our focus is on using caching to proactively fetch data based on mobility prediction, our framework can account for content popularity, which is a factor considered in traditional caching and data placement. Moreover, although our objective is to reduce the data transfer delay, the objective can include other forms of cost such as network bandwidth or monetary.

A novel aspect of the proposed approach is the use of   congestion pricing  that considers the demand for caching and the available storage  to efficiently utilize limited cache capacity, while reducing the  delay  for transferring the requested data to a mobile.
Congestion pricing also allows us to solve the proactive caching problem in a distributed manner, by deciding independently for each data object where it should be cached.
Congestion pricing has been extensively studied for network flow and congestion control.
To the best of our knowledge, the present work is the first to apply it to proactive caching.

The problem of selecting caches to proactively fetch data objects requested by mobiles is fundamentally different from both  conventional caching  and  data (or content) placement. Both these problems involve caching data objects closer to potential requesters, based on the popularity of the objects. In contrast, the  problem investigated in this paper involves proactively fetching data objects  requested by mobiles based on their mobility; different mobiles, independent of the  objects they request, can have different mobility patterns, which can result in different proactive caching decisions. A second difference is that, when a mobile eventually moves to a new  attachment point, the cache space that was occupied by the data requested by the mobile is freed.
Also, proactive caching involves selected caches pulling data from sources and storing it for a short-term, whereas data placement involves pushing data to caches and storing it for a longer-term.
Finally, our proactive caching approach  does not perform eviction or replacement when a cache is full, as in conventional caching; rather, our approach uses congestion pricing to ensure that the objects proactively cached are those for which the highest delay gains are achieved.
Interestingly, prior work has found that  hierarchical or cooperative caching isn't helpful when the user population is above some relatively small threshold \cite{Wol++99}, which is due to the heavy-tailed object distributions. However, such results are not applicable to the two-level proactive cache hierarchy discussed in this paper, where the decision to cache data in  leaf and mid-level caches depends on the mobility patterns and the gains from reduced transfer delay.

Proactive caching of data objects requested by  a mobile  in caches (or proxies) close to the mobile's future  attachment points requires knowledge or prediction of  mobile user requests.
Knowledge of user requests is available \emph{natively}  in Information-Centric Networking (ICN) architectures, which employ a receiver-driven model where receivers request content by its name from one or more publishers, thus supporting receiver mobility \cite{Xyl++12}.
Alternatively, one can predict the data that a mobile is likely to request \cite{Hig++12,Lym++12}.
Note that the focus of this paper is not on how mobility prediction can be performed, nor how the data objects requested are known or predicted. Rather, the paper focuses on developing and evaluating efficient schemes that exploit knowledge or prediction of mobility and data requests  to proactively cache data  to reduce the transfer delay in mobility scenarios.

In summary, our contributions  are the following:
\begin{itemize}
\item We propose a distributed proactive caching approach that selects caches to proactively fetch data objects based on mobility information and  uses  congestion pricing to efficiently utilize cache storage.
\item The proposed approach can be  applied to the case where data objects have different sizes and to a two-level hierarchy of caches.
\item Our modeling framework considers both the case where the delay does not depend on the size of the requested data objects  and the case where the delay is a function of the size of the requested data objects.
\item We present a comprehensive set of evaluation results that show how various system parameters influence the delay gains of  the proposed approach, which achieves robust and good performance relative to  an oracle and an optimal scheme for  a flat cache structure.
\end{itemize}
The rest of the paper is structured as follows: In Section~\ref{sec:related} we discuss related work, identifying where it differs from the work in this paper. In Section~\ref{sec:epc} we present the  models and procedures for efficient proactive caching. Specifically, in Section~\ref{sec:flat} we consider  a flat cache structure while in Section~\ref{sec:two-level} we consider a two-level cache hierarchy; for both, we assume that  delays are independent of the requested object sizes. The case where  delays are a function of the requested  object sizes is discussed in Section~\ref{sec:utility}.
In Section~\ref{sec:evaluation} we evaluate the proposed  proactive caching scheme for scenarios where the delays experienced by all mobiles are the same and for scenarios that involve a scaled-down Internet topology, where different mobiles can experience different delays for transferring an object from its source. Finally, in Section~\ref{sec:conclusions} we conclude the paper identifying directions for future work.

\vspace{-0.09in}
\section{Related work}
\label{sec:related}
\vspace{-0.03in}

\mynote{
\begin{itemize}
\item Identify why problem is different from replication and traditional caching.
\item ICN: knowledge of requested data
\item Proactive caching/fetching: \cite{Gol++12,Mal++12,gaddah2010extending} ...
\item Include application-oriented flow control \cite{Wang++06} as related work, since we use its results in the model where the delay depends on the object size. Perhaps not.
\end{itemize}

}

In the context of publish-subscribe ICN networks, work related to proactive caching includes \cite{burcea2004disconnected,gaddah2010extending}.
The work in \cite{burcea2004disconnected} considers proactive caching based on prediction, but does not investigate  specific algorithms.
The work in \cite{gaddah2010extending} proposes proactive caching in all caches that lie one-hop ahead, close to all network attachment points the mobile can attach to; such a solution will of course lead to inefficient utilization of caches close to attachment points that a mobile has a small probability to attach to. On the other hand, our approach involves selecting a subset of caches close to future network attachment points, based on the probabilities that the mobile  moves to these points.
Moveover, our investigations compare the proposed proactive caching approach with a naive approach that caches data objects in all one-hop caches, subject to the availability of cache storage.
The  work in this paper differs from our previous work on proactive caching in ICN networks \cite{Vas++12}, in that (1) the current paper proposes an approach based on congestion pricing to efficiently utilize cache storage, (2) we propose a proactive caching approach for a two-level cache hierarchy, and (3a) we consider the case where the delay is independent of the requested data object size and (3b) the case where the delay is a function of the object size.

The feasibility of using prediction together with prefetching for vehicular WiFi access is investigated in \cite{Des++09}, which  develops  a prefetching protocol, but does not propose or evaluate specific prefetching algorithms.
The work of \cite{Gol++12} investigates  proactive caching of video content in caches located in femtocells.
Although the objective of \cite{Gol++12} is similar to our objective, there are key differences in both the network model and the proposed solution:
First, \cite{Gol++12} considers a network of partially overlapping femtocells, which differs from the  model considered in this paper.
Second, we consider proactive caching of data requested by each mobile, based on the probability the mobile will move to different future attachment points; once the mobile moves, the data can be removed from all caches  it was proactively fetched. On the other hand, the problem of \cite{Gol++12} is essentially a data placement problem, which considers the popularity of the requested data \cite{App++10}. Finally, we consider a decentralized approach using   congestion pricing  to efficiently utilize cache storage. On the other hand, \cite{Gol++12} proposes a centralized greedy algorithm.
\mynote{Check if it is worth writing something like the following:
On the other hand, our proposed approach achieves performance that is very close, and in some cases coincides with that of the optimal, for the case of a flat cache structure, and has performance which is very close to an oracle.
}
The work of \cite{Mal++12} considers proactive caching (or seeding) to minimize the peak of the total load in a cellular network, subject to user impatience constraints, which essentially impose a maximum delay for the content to be available to users requesting it. An offline water-filling algorithm that determines the content transfer schedules, initially assuming perfect knowledge of future user interests, is proposed; the algorithm is later adapted to handle uncertainty of user interests and traffic.
Our work differs in that our constraint is the cache storage, rather than the network capacity as in \cite{Mal++12}. Furthermore,  \cite{Gol++12,Mal++12} do not consider the case where the transfer delay is a function of the object size.

\vspace{-0.09in}
\section{Efficient proactive caching}
\label{sec:epc}
\vspace{-0.03in}

Proactive caching is used to prefetch data requested by a mobile, so that it is immediately available when the mobile connects to its new network attachment point, thus reducing the  transfer delay.
We consider two cases for the transfer delay. In the first case, the transfer delay is independent of the size of the requested data objects, while in the second the transfer delay is a function of the data object size.

The transfer delay is independent of the requested data object size when the object size is small, e.g. fire/security alerts, thus the transfer delay is determined primarily by the propagation delay rather than the transmission delay.
The delay for obtaining an object from its the original remote source is  denoted $D_{\R}$, whereas the delay  for obtaining an object from the local cache is denoted $D_{\Leaf}$.
These delays  can depend on the distance to the source or cache, e.g. in number of hops.
Note that, although we use the term delay, $D_{\R}$ and $D_{\Leaf}$
can in general include the cost (e.g. network, monetary) for obtaining data from the source or a remote location, which is independent of the requested data object size.
For the above assumption of the transfer delay,
Section~\ref{sec:flat} describes the proactive caching approach for a flat set of caches, while Section~\ref{sec:two-level} considers proactive caching in a two-level cache hierarchy.

The transfer delay will depend on the data object size when it is dominated by the transmission delay, e.g.  for large  objects such as video files. For example, when the mobile is connected to a WiFi hotspot (or 3G/4G small cell), the delay for transferring a data object from a local cache is proportional to the object size and inversely proportional to the  WiFi (or 3G/4G) data rate.
On the other hand, if the  object is transferred from a remote location over a lower capacity  backhaul link (e.g. xDSL) that connects the hotspot (or  small cell) to the Internet, then the delay
is inversely proportional to the backhaul rate, which is smaller than the WiFi (or 3G/4G) rate. As above, the delay can actually involve any cost that is proportional to the object size, e.g. cost per unit of transferred data in the case of data volume charging.
Section~\ref{sec:utility} describes our proactive caching approach when the  delay is a function of the requested object size.

\vspace{-0.10in}
\subsection{Proactive caching in a flat cache structure}
\label{sec:flat}
\vspace{-0.03in}

\mynote{
Key points:
\begin{itemize}
\item Simple decision function for each request/at each cache. This decision function can be applied at mobile device/user or cache. Depending on where the decision is taken, we might need to communicate information on the transition probabilities.
\item Distributed. Decision for each request and each cache can be taken independently.
\item cache price is a congestion price that is updated based on demand. This is a new contribution related to previous proposals.
\item Problem is difficult with different object sizes. Our approach can handle different object sizes.
\item Be careful with use of data object and mobile.
\end{itemize}

}

In this section we present our approach for  selecting the caches that should proactively fetch data objects, in the case of a flat set of caches and when the transfer delay is independent of the size of the requested data objects. Under this assumption, as we will see, the data object is either fully cached or not cached.
Our objective is to minimize the average delay across all requested objects, subject to the cache storage constraints.
Note that in a flat cache structure, the caches are independent, hence prefetching decisions are also independent.

\begin{figure}[b]
\vspace{-0.1in}
\centering
\includegraphics[width=1.8in]{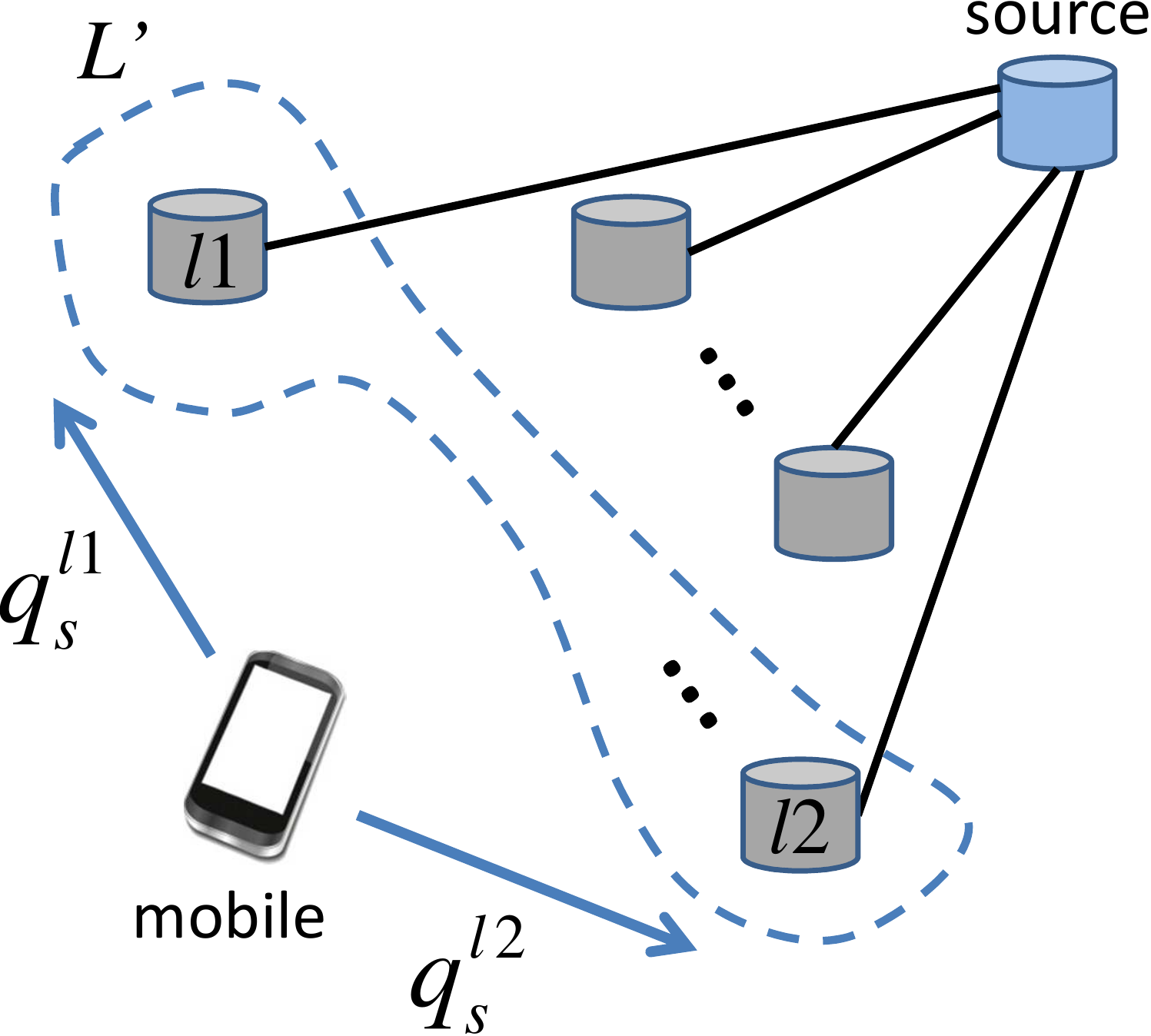}
\vspace{-0.05in}
\caption{The proactive caching problem involves selecting, based on the mobile transition probabilities, the set $L'$ of caches to proactively cache object $s$, in order to achieve (\ref{eq:target}) while satisfying (\ref{eq:constraint}). In the proposed  scheme the decision  to proactively prefetch a data object is taken independently for each cache.}
\label{fig:flat}
\end{figure}

Let $q^l_s$ denote the probability that the mobile requesting object $s$ moves to cache $l$ and $B_l$ denote the maximum storage at cache $l$; we initially assume that all objects have the same size $o$.
Note that the probability $q^l_s$ can depend on the specific mobile, its current or past location, and the time instant, but for simplicity we do not make this dependence explicit in the notation.
Also, let $S_l$ be the set of objects requested by mobiles that have non-zero probability to move to cache $l$ and $L$ be the set of caches. We define the following optimization problem:
\begin{eqnarray}
\min_{b^l_s} &  \displaystyle\sum_{s \in S_l} \mathcal{D}_s \label{eq:target} \\
\, & \nonumber \\
\mbox{subject to} & \displaystyle\sum_{s \in S_l} o \cdot b^l_s \leq B_l \, , \; \; \forall l \in L,\label{eq:constraint}
\end{eqnarray}
where $\mathcal{D}_s=\sum_{l \in L} \mathcal{D}^l_s$ is the average delay for obtaining object $s$ and $b^l_s$ equals one if the object $s$ is proactively fetched in cache $l$ and zero if it is not proactively fetched in cache $l$.
$\mathcal{D}^l_s$ is equal to  $q^l_s D_{\R}$ if the object is not in cache $l$ ($b^l_s=0$) and needs to be obtained from its original remote location and $q^l_s D_{\Leaf}$ if the object is stored in  cache $l$ ($b^l_s=1$).
The above optimization problem involves selecting  for each data object $s$ requested by a mobile, based on the mobile transition probabilities, Figure~\ref{fig:flat}, the subset $L' \in L$ of caches that will proactively fetch object $s$ so that it is immediately available to the mobile when it connects to an attachment point close to the cache, in order to achieve the optimization target (\ref{eq:target})  while satisfying the cache storage constraint (\ref{eq:constraint}).

In order to efficiently utilize the cache storage, we introduce a congestion price $p_l$ which is adapted based on the demand for caching and the available storage.
Specifically,
\begin{equation}
p_l(t+1)= \left [ p_l(t)+ \gamma \left ( o \cdot b^l(t) - B_l \right ) \right ] ^+ \, , \label{eq:cong_price}
\end{equation}
where $b^l(t)$ is the aggregate demand at cache $l$ at time $t$ and $\gamma$ is the price update factor, which  determines how quickly the cache congestion price adapts to changes of the demand for caching.
For the evaluation in Section~\ref{sec:evaluation}, the cache price is updated when a new cache request appears.

\mynote{
The implementation uses a slightly  different cache price update. Need to decide if we need to change above.
}

The decision to proactively fetch an object $s$ at cache $l$ is based on  the following  rule:
\begin{equation}
b^l_s=
\left \{
\begin{array}{cc}
0 \; \;  & \mbox{ if } \; \; q^l_s (D_{\R}-D_{\Leaf}) < p_l  \\
$\,$ & \nonumber \\
1 \; \;  & \mbox{ if } \; \; q^l_s (D_{\R}-D_{\Leaf}) \geq p_l
\end{array}
\right .
\label{eq:decision}
\end{equation}
The above decision rule provides a distributed and decentralized approach to decide, for each object $s$ and for each cache $l$, whether the object should be prefetched.
Of course, the object will actually be prefetched only if cache storage is available.
Adjusting the cache price using (\ref{eq:cong_price}) directs the system towards efficient use of cache storage, while achieving the optimization  target (\ref{eq:target}). Specifically, when the cache is underutilized, i.e. cache storage is available,  the congestion price decreases, thus allowing more objects to be proactively fetched in the cache based on the decision rule (\ref{eq:decision}). On the contrary, when the cache demand  is larger than the cache size, then the price increases which in turn, due to (\ref{eq:decision}), reduces the number of objects that are requested to be  cached.
Furthermore, when the cache price is such that the  amount of requested cache is equal to the cache storage, then due to the decision rule (\ref{eq:decision}) we are certain that these requests correspond to the highest values of $q^l_s (D_{\R}-D_{\Leaf})$, hence the minimum in (\ref{eq:target}) is achieved.
Note that content popularity can be incorporated in the above model if we replace $q_s^l$ with the sum $q_s^l+r_s$, where $r_s$ is the popularity of object $s$.

Two practical issues related to the application of the  decision procedure (\ref{eq:decision}) include where the decision is taken and if the decision is to proactively cache an object, when should prefetching start.
Regarding where the decision is taken, one option is for the mobile requesting object $s$, or some proxy on behalf of the mobile, to inform all  caches located close to its possible future attachment points about its transition probability: When cache $l$ learns the probability $q^l_s$, then together with the delays $D_{\Leaf}, D_{\R}$ and the cache congestion price $p_l$ it can apply the decision rule (\ref{eq:decision}). Alternatively, the  caching decision can be taken at the mobile, or its proxy, in which case it would need to learn the delays $D_{\Leaf}, D_{\R}$ and the cache  price $p_l$ from all caches it has some probability to connect to.
The second issue of when to start to prefetch a data object is related to the time interval after which the mobile connects to its next network attachment point and the time for the cache to download the requested object from its remote location.

When a mobile moves to its new attachment point, then it can directly receive the requested object from the local cache, if the cache had prefetched the requested  object. Otherwise, the mobile  obtains the requested object from the original source. When a mobile connects to its new attachment point, then the space occupied in all caches that proactively  cached that mobile's data can be freed; of course, if the local cache at the mobile's new attachment point had prefetched the mobile's data, then the corresponding space will be freed after the data is transferred to the mobile. The above actions require communication and cooperation among caches.

The  model presented above can be extended to objects with different sizes, by replacing the constraint in (\ref{eq:constraint}) with
$$
\sum_{s \in S_l} o_s \cdot b^l_s  \leq B_l \, ,
$$
where $o_s$ is the size of  object $s$.
Additionally,  the cache price $p_l$ on the right side of the inequalities in (\ref{eq:decision}) should be replaced with $o_s \cdot p_l $.
For objects with different sizes, the optimization problem becomes identical to the \emph{0/1 Knapsack Problem}, which falls within the class of NP-hard problems.

Another extension is when the remote and local delay is different for different data objects and caches, hence they can be denoted $D_{\R}^{s,l}$, $D_{\Leaf}^{s,l}$ and the
decision rule (\ref{eq:decision}) can be adapted accordingly. Instead of maintaining different delays  for different objects or mobiles, delays can be associated with  object or mobile types, or can depend on the mobile's initial network attachment point. In all these cases, the actual values of the delay can be estimated in a measurement-based manner.

\subsubsection{Optimal for equal-size objects}
\label{sec:opt}

\mynote{
Key points:
\begin{itemize}
\item Problem assumes some set of cache requests, and seeks to find which and where to cache objects.
\item Optimal solution is implemented in rounds.
\end{itemize}
}

For a given set of cache requests, the optimal in the case of equal size data objects, can be obtained for a flat set of caches as follows: For each cache $l$, we order the cache requests in  decreasing value of $q^l_s(D_{\R}-D_{\Leaf})$. Then, starting from the request with the highest $q^l_s(D_{\R}-D_{\Leaf})$, we fill the cache until the constraint $B_l$ is reached.

The above procedure for obtaining the optimal is performed in rounds: in the beginning of each round we have a given set of cache requests. This is unlike the solution based on cache congestion pricing, where the decision of whether to cache an object is taken iteratively, based on (\ref{eq:decision}), for each cache request, hence can be applied on-line.
A practical issue with the optimal solution is the duration of each round, which determines the number of cache requests considered; this duration depends on the time interval after which a mobile connects to its new attachment point and $D_{\R}$.

\vspace{-0.10in}
\subsection{Proactive caching in a two-level hierarchy}
\label{sec:two-level}
\vspace{-0.03in}

\mynote{
\begin{itemize}
\item  Show why problem is difficult, prove NP.
\item Separate problem to solution of two problems, which consider a single-level cache set, hence can be solved distributed as in previous subsection. Then, from the two solutions take the one that yields smallest delay.
\item Show communication of information between middle cache and leaf caches. Figure with two-levels of caches.
\item When there are two or more mid-level caches, then each mid-level cache implements same procedure described. Each leaf cache is child of one mid-level cache.
\item possibly add pseudocode algorithm to make approach explicit/clear.
\item Alternative: maintain a congestion price for mid-level cache. Decide to cache object in mid-level cache only if Delay(object fetched in mid-level cache)-Delay(object not fetched in mid-level cache) $\geq$ $p_{\mbox{\tiny mid-level cache}}$.
    \end{itemize}
}

\begin{figure}[b]
\vspace{-0.15in}
\centering
\includegraphics[width=2.4in]{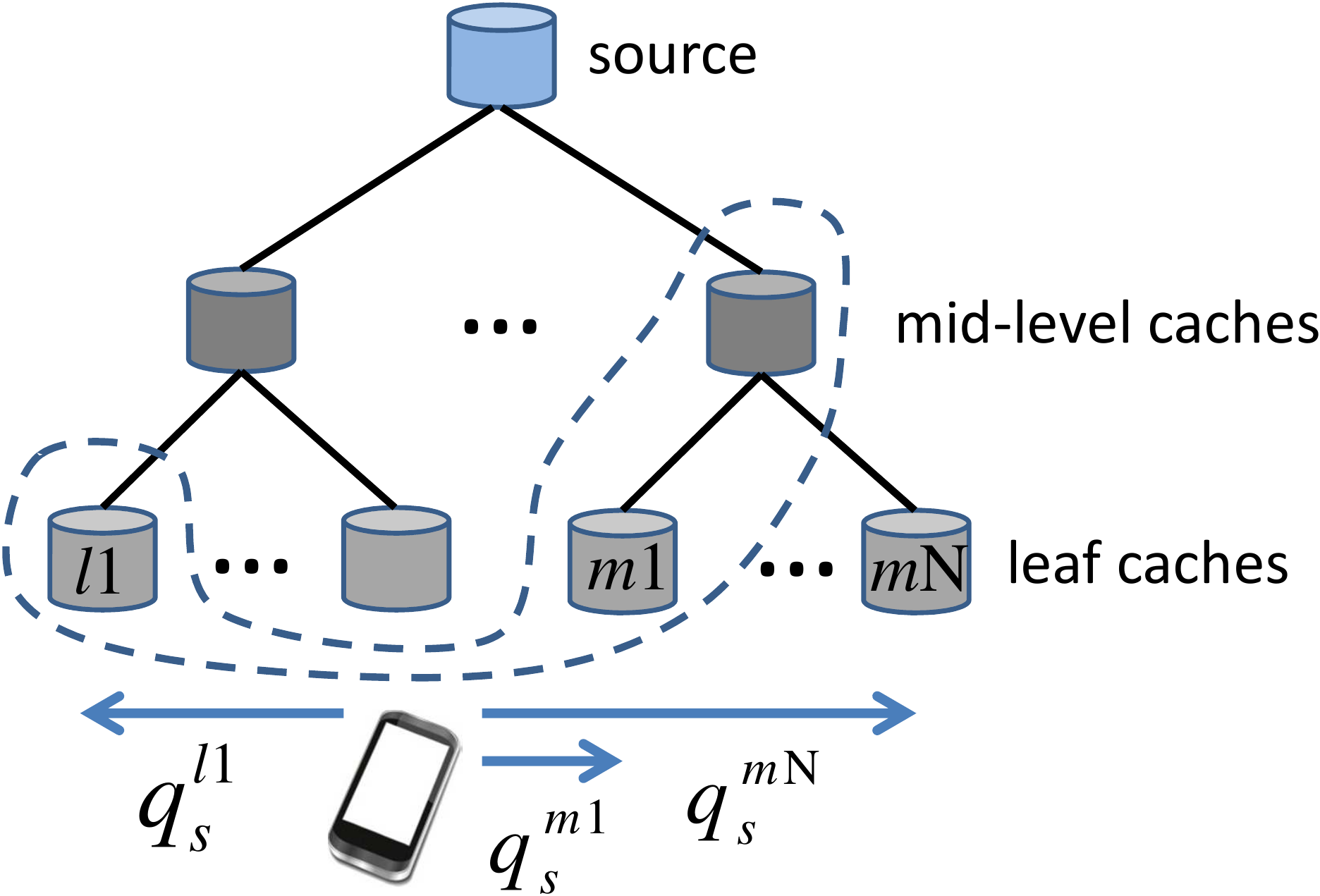}
\caption{In a two-level cache hierarchy, each mid-level cache cooperates with its leaf caches to decide which  will proactively cache a data object. No such cooperation is necessary between mid-level caches.}
\label{fig:2level}
\end{figure}

Next we consider a two-level cache hierarchy. Each leaf node can be under only one mid-level cache, Figure~\ref{fig:2level}, and  an object can be proactively fetched at a leaf cache, at a mid-level cache, or both.
The delay for obtaining an object from a mid-level cache is $D_{\M}$, which satisfies $D_{\Leaf}<D_{\M} < D_{\R}$.
At any time instance, there will be a given set of cache requests from the  mobiles that are active at that time instant. For each such time instance, the proactive cache problem for a two-level cache hierarchy has similarities with the data placement problem \cite{Bae+08}, where the probability of an object being requested at a specific cache is given by the probability of the mobile moving to the corresponding network attachment point. The authors of  \cite{Bae+08} show that the data placement problem with different object sizes is NP-complete. Although there are cases where the placement problem in a hierarchical network with equal size objects can be solved in polynomial time \cite{Kor+01}, such solutions have a high polynomial degree  and apply to an offline version of the problem.


In  a two-level cache hierarchy, the leaf cache can correspond to caching that is performed at a local area network, such as femto/small cell, hotspot or home network, whereas the mid-level cache can correspond to caching at the ISP that connects the local network to the Internet.

\mynotex{ADD MORE DETAILS in figure caption}

Our approach to solve the proactive caching problem  in a two-level cache hierarchy first considers two flat cache selection problems: one assuming that the object is proactively fetched in the mid-level cache and the other assuming that the object is not proactively fetched in the mid-level cache.
Each of the aforementioned flat cache problems can be solved using the distributed approach presented in Section~\ref{sec:flat},
by having the mid-level cache send the leaf caches  the delay  $D_{\R}$, which is the delay for obtaining the object from the remote source, and the delay $D_{\M}$, which is the delay for obtaining the object from the mid-level cache.
Each leaf cache, using (\ref{eq:decision}), decides whether to cache the specific data object for each of the two problems:
For the problem where the object is assumed not to be cached in the mid-level cache, the leaf cache uses (\ref{eq:decision}) to decide if the object should be prefetched in the leaf cache. Whereas, for the problem where the object is assumed to be cached in the mid-level buffer, the leaf cache uses (\ref{eq:decision}) replacing $D_{\R}$ with $D_{\M}$ to decide if the object should be prefetched in the leaf cache.
Each leaf cache informs the mid-level cache about the resulting average object delay for each of the two problems: $q^l_s D_{\R}$ or $q^l_s D_{\M}$ if the decision is not to proactively fetch the object  and $q^l_s D_{\Leaf}$ if the decision is to proactively fetch the object in the leaf cache.

After receiving from all leaf caches the  delay for the two problems, the mid-level cache takes the sum of the delays for each problem: $\mathcal{D}^{\mmid}_{\M}$ in the case the mid-level cache  proactively fetches the data object and $\mathcal{D}^{\mmid}_{\R}$ in the case the mid-level cache does not proactively fetch the data object. The decision of whether to cache or not cache a data object in the mid-level cache is determined based on a decision rule that resembles (\ref{eq:decision}):
If $\mathcal{D}^{\mmid}_{\R} - \mathcal{D}^{\mmid}_{\M} \geq p_{\mmid}$ then the object is proactively fetched in the mid-level cache, otherwise it is not fetched. The variable  $p_{\mmid}$ is the congestion price for the mid-level cache, which is updated in a similar manner as the congestion price for the leaf caches (\ref{eq:cong_price}), but based on the demand for caching in the mid-level cache and the storage available in the mid-level cache.
Following its decision, the mid-level cache then informs the leaf caches which delay factor ($D_{\R}$ or $D_{\M}$) they should use in (\ref{eq:decision}) to eventually decide whether to cache a data object.

The above procedure is distributed and requires some cooperation between the mid-level cache and its leaf caches.
Moreover, it can be applied to a hierarchy with more than one mid-level caches, as long as each leaf cache is a child of only one mid-level cache. The proactive caching decision for each mid-level cache and its leaf caches follows the above approach, and can be performed independently of other mid-level caches.

\vspace{-0.09in}
\subsection{Proactive caching when delay is a function of object size}
\label{sec:utility}
\vspace{-0.03in}

\mynote{
\begin{itemize}
\item  Extension of this model to a two-level hierarchy is not straightforward.
\item Perhaps more important, a two-level hierarchy of this model may not correspond to a scenario of practical interest: In a hotspot/small cell topology, the bottleneck is the hotspot/small cell backhaul. To have two-levels, we would need to have a link that is between the WiFi/radio link and the backhaul which has rate $R_{\bkhl} < R_{\M} < R_{\wifi}$. In practical systems, such a link does not exist.
\item An alternative is to write the utility as a function of the average delay, $\mathcal{U}_s(\sum_l q^l_s \mathcal{D}_s(x^l_s))$. Indeed, one can argue that such a form is more natural. However, such a form makes the utility $U_s()$ a function of variables $x^l_s, l\in L$, each of which encounter  a different congestion price. This problem is similar to  multipath flow control optimization models. A disadvantage is that it leads to a more complex decision rule compared to (\ref{eq:decision2}).
\item Addition of this model is to make framework more complete. Evaluation results for which model will/can not be included.
\end{itemize}
}

In Sections~\ref{sec:flat} and \ref{sec:two-level} we assumed that the transfer delays $D_{\Leaf},D_{\M}, D_{\R}$  are independent of the data object size.
In this section we consider the case where the transfer delay is a function of the data object size. As we will see, in this case there can be gains if a part of an object, and not necessarily the whole object, is proactively cached.

Let $R_{\Leaf}$ be the  rate for transferring to a mobile  data  stored locally at a cache close to the mobile's  attachment point, and $R_{\R}$ be the rate for transferring to a mobile  data from a  remote source. We assume that $R_{\R}<R_{\Leaf}$, which justifies why proactively caching can be beneficial.  As an example, $R_{\Leaf}$ can be the rate for transferring data across a WiFi or 3G/4G interface, while $R_{\R}$ can be the rate for transferring data across a hotspot or small cell's backhaul link, which is typically smaller than the rate of a WiFi or 3G/4G interface.

Assume that some part $x^l_s$ of object $s$ that has size $o_s$ is proactively fetched to  cache  $l$, whereas the remaining part $o_s-x^l_s$ would need to be obtained from the object's original remote location. If  the mobile requesting object $s$ moves to an attachment point close to cache $l$, then the delay for transferring the whole object $s$ is given by
\begin{equation}
\mathcal{D}_s(x^l_s) =  \frac{x^l_s}{R_{\Leaf}} + \frac{o_s-x^l_s}{R_{\R}}  =  \frac{o_s}{R_{\R}} - \left ( \frac{1}{R_{\R}} - \frac{1}{R_{\Leaf}}  \right ) x^l_s  \, .
\label{eq:delay}
\end{equation}
In the last equation we have assumed that the rates $R_{\Leaf}, R_{\R}$ are the same for all caches $l$. The results presented below are similar when the rates are different for different caches.

\begin{figure}[tb]
\begin{center}
\begin{tabular}{c}

\begin{minipage}[b]{0.5\linewidth}
\centering
\hspace{-0.22in}
\includegraphics[width=1.4in] {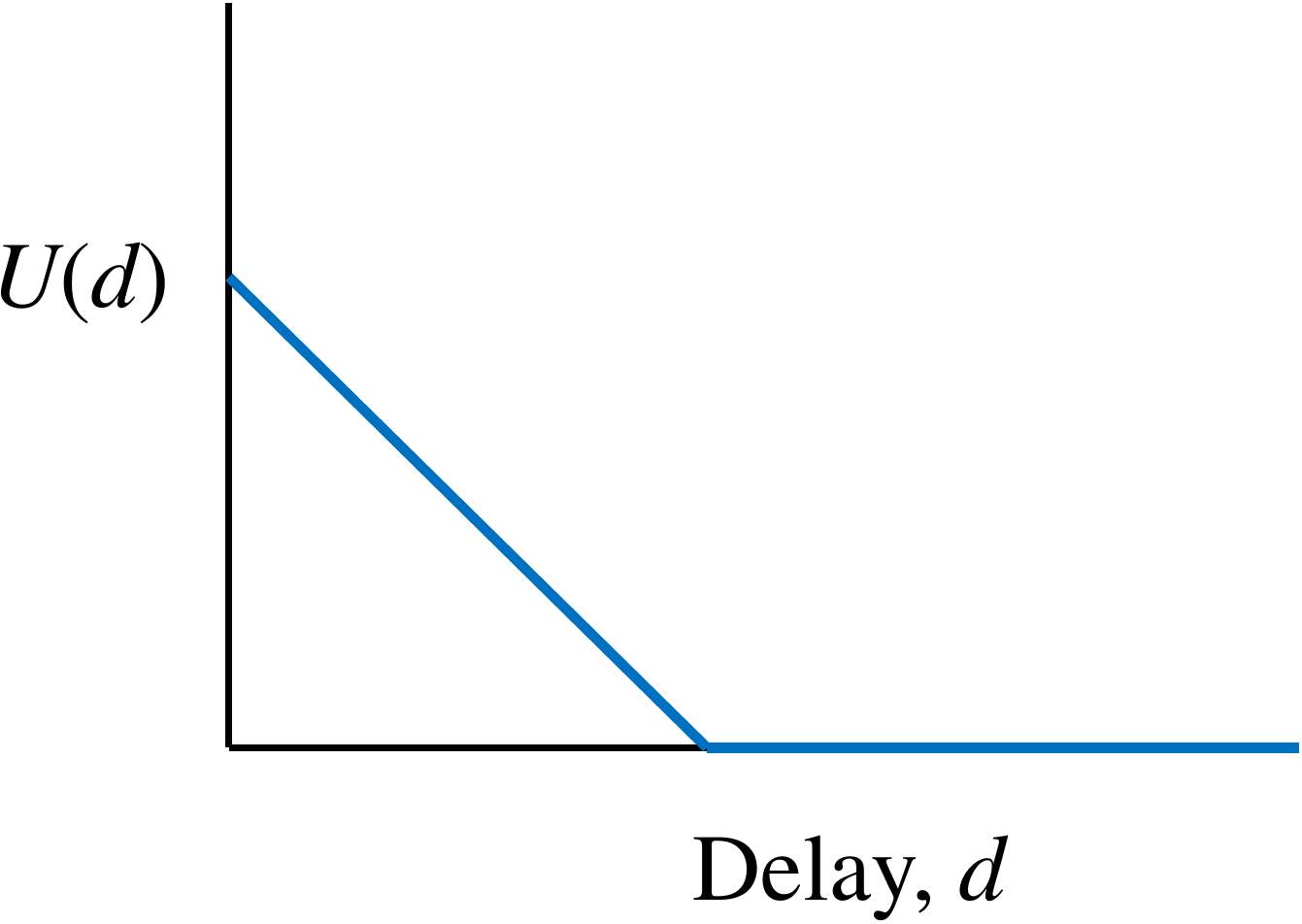}\\
{\footnotesize {(a) }}
\end{minipage}

\begin{minipage}[b]{0.5\linewidth}
\centering
\hspace{-0.22in}
\includegraphics[width=1.4in]{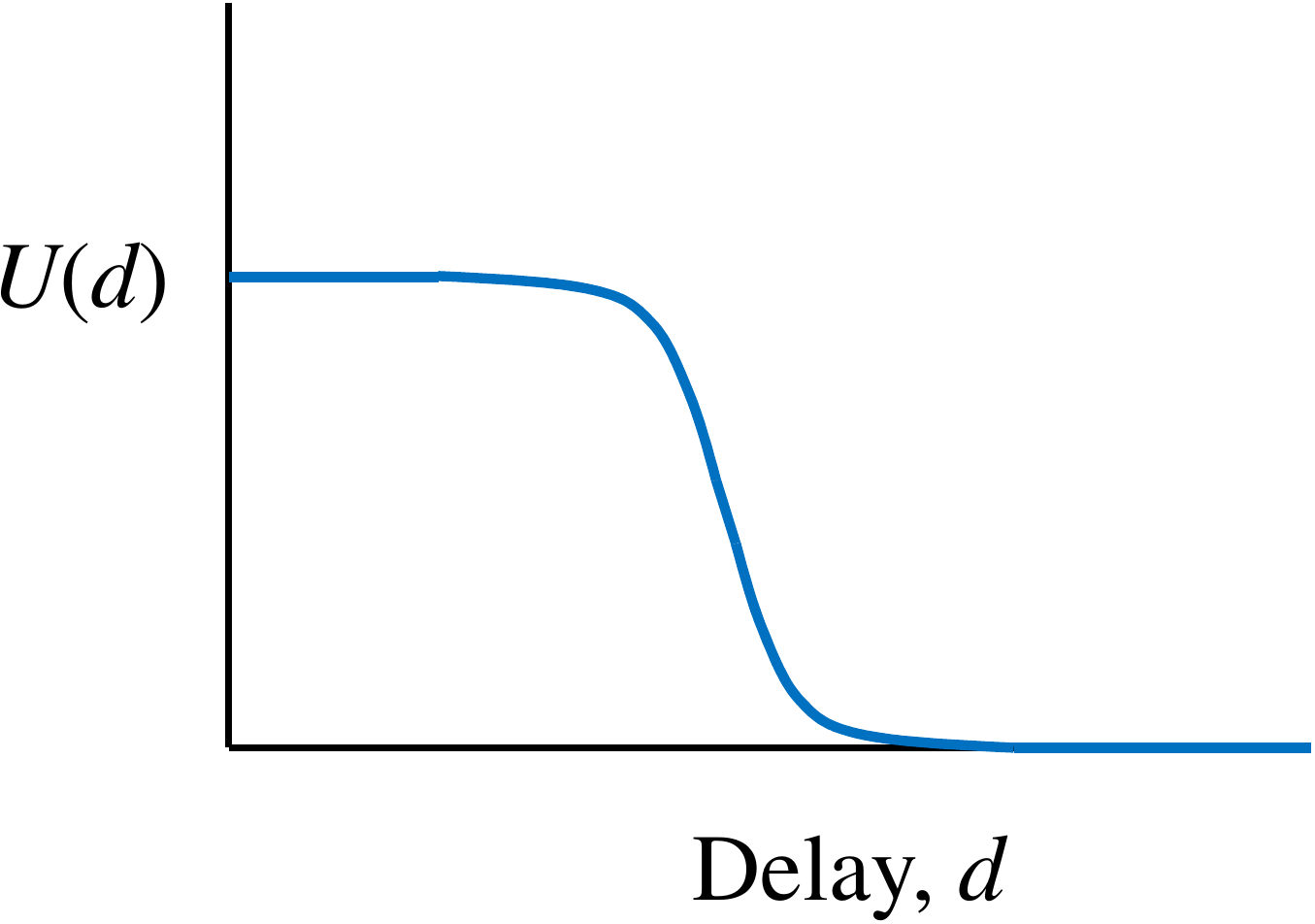}\\
{\footnotesize  {(b) }}
\end{minipage}\

\end{tabular}
\end{center}
\vspace{-.12 in}
\caption[]{\protect Example utilities as a function of delay.}
\label{fig:utilities}
\vspace{-0.17in}
\end{figure}

Consider a utility function $\mathcal{U}_s(d)$ that represents a mobile user's valuation for delay $d$ in transferring object $s$. $\mathcal{U}_s(d)$ is a decreasing function, and possible shapes for it are shown in Figure~\ref{fig:utilities}. Figure~\ref{fig:utilities}(a) corresponds to the case where a user obtains no value when the delay is above some maximum threshold, and obtains  a value that increases linearly as the delay approaches zero.
The sigmoid utility in Figure~\ref{fig:utilities}(b) corresponds to the case where a user obtains maximum value when the delay is below some minimum threshold, while he obtains zero value when the delay is above some maximum threshold; such a curve approximates the exact step utility of hard real-time applications, with strict delay requirements.

We can define the utility $U^l_s(x^l_s)=\mathcal{U}_s(\mathcal{D}_s(x^l_s)/q^l_s)$, which is now a function of the part $x^l_s$ of  object $s$ that is proactively fetched in cache $l$.
Note that in the last equation we have added the factor $1/q^l_s$ to account for the transition probability to cache $l$. Hence, if a mobile has a higher probability to move to cache $l$ compared to some other cache, then it would need to proactively cache a larger amount of the data object to achieve the same utility.
We assume that the utility function  $U^l_s(x^l_s)$ is continuous and strictly increasing  in the interval $[m^l_s,M^l_s]$, where $m^l_s \geq 0$ and $m^l_s < M^l_s \leq o_s$ are minimum and maximum values of $x^l_s$ for which the following hold:
$U^l_s(x^l_s)=U^l_s(m^l_s)$ for $x^l_s \leq m^l_s$ and $U^l_s(x^l_s)=U^l_s(M^l_s)$ for $x^l_s\geq M^l_s$.
As an example, a utility function that corresponds to Figure~\ref{fig:utilities}(a) is $U^l_s(x^l_s) = \frac{\mathcal{R}}{q^l_s} ( x^l_s - m^l_s )$, for $x^l_s \in [ m^l_s,M^l_s ]$, where $\mathcal{R}= \frac{1}{R_{\R}}-\frac{1}{R_{\Leaf}}$.

As in the previous two subsections, we define a cache congestion price that is updated as in (\ref{eq:cong_price}), with $o \cdot b^l(t)$ replaced by $x^l(t)=\sum_{s \in S_l} x^l_s(t)$, which gives the aggregate demand for cache storage at $l$, when data objects can be partially cached:
\begin{equation}
p_l(t+1)= \left [ p_l(t)+ \gamma \left ( x^l(t) - B_l \right ) \right ] ^+ \, , \label{eq:cong_price2}
\end{equation}
where as before $\gamma$ is a price update factor.

The framework below follows the model of \cite{Wang++06}.
We define the following decision rule for selecting the amount $x^l_s$ of object $s$ to be proactively fetched in cache $l$:
\begin{equation}
x^l_s=
\left \{
\begin{array}{cl}
m^l_s \; \; & \mbox{ if } \; \; \frac{1}{U^l_s(m^l_s)} \leq p_l \\
{U^l_s}^{-1} \left ( \frac{1}{p_l} \right ) \; \; & \mbox{ if } \; \; \frac{1}{U^l_s(M^l_s)} < p_l < \frac{1}{U^l_s(m^l_s)} \\
M^l_s \; \; & \mbox{ if } \; \; p_l  \leq \frac{1}{U^l_s(M^l_s)}
\end{array}
\right .
\label{eq:decision2}
\end{equation}
Unlike the  binary decision rule (\ref{eq:decision}), where a data object is either fully cached or is not cached, with (\ref{eq:decision2}) it may happen that only a part of object $s$ is fetched at cache $l$: $x^l_s$ is a continuous variable with values in  $[m^l_s,M^l_s]$, where $m^l_s \leq 0$ and $m^l_s< M^l_s \leq o_s$.

Using the results from \cite{Wang++06}, it can be shown that a system where
the amount  $x^l_s$ of object $s$ to be proactively fetched in cache $l$ is determined by (\ref{eq:decision2}) and the cache  price is updated according to  (\ref{eq:cong_price2}), with an appropriately small value $\gamma$,  will converge and solve the  following optimization problem:
$$
\begin{array}{cc}
\displaystyle\max_{m^l_s \leq x^l_s \leq M^l_s} & \displaystyle \sum_{l\in L} \sum_{s\in S} V_s(x^l_s) \nonumber \\
$\,$ & \nonumber \\
\mbox{subject to} & \displaystyle\sum_{s \in S_l} x^l_s \leq B_l \, , \; \; l \in L \nonumber
\end{array}
$$
where $S$ is the set of all data objects, $S_l$ is the set of objects cached at $l$, $L$ is the set of all caches, and
$$
V_s(x^l_s) = \int_{m^l_s}^{x^l_s} \frac{1}{U^l_s(y)} dy \, , \; \; m^l_s \leq x^l_s \leq M^l_s \, .
$$
As an example, for the utility $U^l_s(x^l_s) = \frac{\mathcal{R}}{q^l_s} ( x^l_s - m^l_s )$, we have $V_s(x^l_s)= \frac{q^l_s}{\mathcal{R}} \log (x^l_s-m^l_s)$.

Moreover, at the optimum the requested amount of data to be proactively cached $\{ {x^l_s}^*: s \in S, l \in L \}$ achieves utility proportional fairness: for any other set of cached data sizes $\{ x^l_s: m_s^l \leq x^l_s \leq M^l_s, s \in S, l \in L \}$,
$$
\sum_{l \in L} \sum_{s \in S} \frac{x^l_s - {x^l_s}^*}{U^l_s({x^l_s}^*)} \leq 0 \, .
$$
Utility fairness can be seen as a way to allocate limited resources in a manner that is fair from an application perspective, since it takes into account the actual valuation (utility) for a specific amount of resources (cache storage in our case). In contrast, resource-oriented fairness definitions, such as proportional fairness and max-min fairness, seek to allocate resources in a fair manner from a resource-centric perspective, which does not necessarily reflect the requirements at the application level. Thus, considering utility fairness in the model of this section results in an application-oriented approach for performing proactive caching.

\vspace{-0.09in}
\section{Evaluation}
\label{sec:evaluation}
\vspace{-0.03in}

\mynotex{Comparison with/influence of
\begin{itemize}
\item Optimal scheme for single-level and oracle
\item distribution of a given/fixed total cache storage to leaf and mid-level caches, for different total cache storage
\item ratio (total cache)/demand, for different values of mid-level buffer
\item influence of mobile transition probabilities
\item influence of delay costs $D_{\Leaf},D_{\M},D_{\R}$
\item probably not: mid-level cache and leaf cache size
\end{itemize}
}

In this section we evaluate the proposed Efficient Proactive Caching (EPC) scheme  using the OMNeT++ simulation framework. We present results for a flat and a two-level cache hierarchy, for data objects that have the same size and when the delay is independent of the object size. Evaluation results for different object sizes and when the delay is a function of the  object size will be included in a followup of this paper.

\vspace{-0.06in}
\subsection{Simulation model}
\vspace{-0.03in}

We consider scenarios where the delay for obtaining a data object is the same for all mobiles (referred to as fixed delay scenarios) and scenarios where the delay for obtaining a data object from the source is variable and depends on the number of hops between the source and the mobile  (receiver); the latter scenario involves a scaled-down Internet topology containing 400~nodes,  with each node representing an autonomous system (AS) \cite{dimitropoulos}.
The value of various system parameters considered in the simulations are shown in Table~\ref{tab:values}.
In the fixed delay scenarios, there are a total of  160~mobile users that issue requests for objects of the same size. The mobiles can move to 8~different network attachment points, where there is a corresponding local leaf cache. At a higher level, there is a mid-level cache. Whenever a mobile performs a handover, we assume that a new mobile enters the system, thus the total number of active mobiles in the system always remains 160.
In the  scenarios with scaled-down Internet topology, we conduct simulations over a set of 10~different neighborhoods, with each neighborhood having 160~mobile users, yielding a total of 1600~mobile users.
Each neighborhood is composed by 8 ASes lying on the edge of the topology. The vast majority of such ASes are stub (access) networks to the Internet. For each neighborhood, we randomly select an initial stub AS and then form the neighborhood by selecting 8 stub ASes based on minimum hop distance from the initial stub AS. The underlying idea is that the selected nodes are neighbors of the initial stub AS, thus typically they should be a few hops away for mobile users  hosted by the initial stub AS. Note that the neighbors are selected so there are no overlaps between different neighborhoods.

A mobile user can move with some transition probability to one of the 8~different attachment points, each with its own local cache. In the fixed delay scenarios there is one mid-level cache, whereas in the scaled-down Internet topology scenarios there is one mid-level cache in each neighborhood.
We consider 4~different sets of mobile transition probabilities, Table~\ref{tab:values}, with different \emph{skewness}, where a higher \emph{skewness} corresponds to a higher probability to move to a particular attachment point (equivalently, cache).
We also assume that the destination network attachment point (equivalently, the destination cache) with the highest transition probability is different for different mobiles, such that the number of active mobiles moving to a specific cache is on average the same, and equal to 20 throughout the simulation.
Finally, note that with the EPC and optimal schemes the transition probabilities used in the caching decisions are measured as the  simulation progresses.

The performance of the EPC scheme depends only on the ratio of delays $D_{\R}/D_{\Leaf}$ and $D_{\M}/D_{\Leaf}$, because the decision rule for both the leaf cache (\ref{eq:decision}) and the mid-level cache has a linear dependence on the delays and the congestion price is adjusted to achieve high utilization; hence we consider the delay ratios rather than the absolute delay values.
Assuming that the leaf cache performs caching at a local area network (e.g.  femto/small cell or hotspot) and that the mid-level cache performs caching at the ISP that connects the local network to the Internet, we have taken  $D_{\M}/D_{\Leaf}=2$ and $5$, which can be seen as the number of hops or the actual delay for obtaining a data object from an ISP cache relative to a local network cache.
We have considered $D_{\R}/D_{\Leaf}= 10$ and 18; together with the values of  $D_{\M}/D_{\Leaf}$, these give values of $D_{\R}/D_{\M}$ in the range $[ 2, 5]$; the lowest value $D_{\R}/D_{\M}=2$ corresponds to the case where an ISP has a direct peering link with the content provider network (source for a data object), in which case their distance is two AS hops.
At the other side, studies have shown that the average inter-AS path length has remained practically constant and equal to 4.2 over the last 12 years \cite{Dha+11}, and for this reason we have selected the highest value of $D_{\R}/D_{\M}$ to be 5. For the scaled-down Internet topology scenarios, we consider $D_{\M}/D_{\Leaf}=5$ and $D_{\R}/D_{\Leaf}=\frac{9}{5} (\mbox{\# hops}-1)+1$; the latter gives an average  $D_{\R}/D_{\Leaf}$ equal to 10, since the average number of hops between a source and receiver in the scaled-down Internet topology is 6.

\mynotex{why we take ratio of delays}

\mynote{XV: how do we define neighborhood}

The performance of the proposed EPC scheme is compared to the optimal scheme described in Section~\ref{sec:opt},  to a naive scheme, and an oracle.
It is important to note that the optimal scheme in the case of a flat cache structure is implemented such that the cache allocation is performed whenever cache storage is freed. In addition to being time consuming, the ability to implement frequent cache allocations can be constrained by the time for actually transferring the data objects to the caches where they are proactively fetched.

With the naive scheme, a mobile requests caching for all data objects, provided that storage is available. With the oracle, for each new cache request we assume that the data object is prefetched (provided there is cache space) by the cache located at  the attachment point where the mobile will eventually connect to (hence the name oracle). Unlike the optimal scheme, which allocates cache storage in rounds considering the requests from active mobiles,  the EPC, naive and oracle iteratively, for each new cache request, take caching decisions that do not change until the corresponding handoff is performed.

\mynotex{\begin{itemize}
\item fixed delay: 10 runs for fixed delay.
\item Internet topology: 10 runs for each neighborhood.
\item each run has duration 10000 handoffs
\item 95\% confidence interval, less than 5\% for all results except 30\% variation of transition probabilities.
\item naive and oracle influences very little by transition probabilities, hence we do not show different results.
\item how long is each run? how many runs do we take for average. conf interval?
\end{itemize}
}

\begin{table}[t]
\caption{Parameter values. The values designated with * are default values, i.e. the values if the specific evaluation scenario does not indicate otherwise.}
    \label{tab:values}
    \centering
 {\scriptsize
        \begin{tabular}{|c|c|c|}
        \hline
        Parameter  &  \multicolumn{2}{c|}{Values} \\ \cline{2-3}
                   &  Fixed delay & Scaled-down Internet topology\\
        \hline \hline
   \# of active mobiles      &  160 & 160 per neighborhood \\
                             &      &  10 neighborhoods \\ \hline
   \# of attachment points &  8  & 8 per neighborhood \\ \hline
   Avg. mobile trans probs & \multicolumn{2}{c|}{SKD50\%: $50\%,20\%,10\%,7.5\%,5\%, 3 \times 2.5\%$} \\
     & \multicolumn{2}{c|}{SKD70\%$^*$: $70\%,2 \times 10\%,3 \times 2.5\%,2 \times 1.25\%$} \\
     & \multicolumn{2}{c|}{SKD90\%: $90\%,3 \times 2\%,4 \times 1\%$} \\ \hline
  Std. dev. of trans probs & \multicolumn{2}{c|}{$5\%^*, 30\%$} \\ \hline
  Delay &  $D_{\M}/D_{\Leaf}=2,5^*$  & $D_{\M}/D_{\Leaf}=5$ \\
        & $D_{\R}/D_{\Leaf}=10^*,18$ & $D_{\R}/D_{\Leaf}=\frac{9}{5}$(\#hops-1)+1 \\ \hline
  Total cache (leaf+mid) & \multicolumn{2}{c|}{$0-320$ objects, default:$240^*$} \\
        \hline
        \end{tabular}
        }
\vspace{-0.05in}
\end{table}

\mynote{TODO:
\begin{itemize}
\item XV: actual value for X.X.
\end{itemize}
}

\mynotex{TODO:
\begin{itemize}
\item Possibly add reference about number of hops, to motivate the ratios $D_{\R}/D_{\Leaf}, D_{\M}/D_{\Leaf}$
\end{itemize}
}

\vspace{-0.05in}
\subsection{Evaluation results}
\vspace{-0.05in}

In this section we compare the various schemes in terms of the gains in reducing the average delay for the mobiles to obtain the requested data objects, compared to the delay if no caching is used, i.e. when all data objects are obtained from the original sources.
Because the performance of the naive and oracle schemes showed very small dependence on the mobile transition probabilities, for these schemes we do not show results for different transition probabilities.

The results shown  are the average of 10 runs for each scenario in the fixed delay case and 10 runs for each neighborhood in the case of the scaled-down Internet topology. Each run has a duration that corresponds to 10.000 handoffs. For these parameters, the 95\% confidence interval was within 5\% of the average values,  hence they are not shown in  the graphs.

\mynote{XV: Check stand. dev. for variable delay.}

The  conclusions from the evaluation are the following:
\begin{itemize}
\item The gains of EPC  are higher when there is more mobility information, i.e. for a higher skewness of the mobile transition probabilities; these gains are close to those of the optimal scheme for a flat cache structure and the oracle.
\item For a higher skewness of the mobile transition probabilities,  EPC achieves higher gains when  more storage is allocated to  leaf caches. Moreover, the gains of  EPC are significantly higher than the naive scheme when more storage is allocated to the leaf caches.
\item Even for a relatively high variation of the mobile transition probabilities, EPC's gains are robust, and  significantly higher  than the naive scheme, when the  skewness of transition probabilities is high and when more storage is allocated to leaf caches.
\end{itemize}

\subsubsection{Comparison of Efficient Proactive Caching (EPC)  with the naive, optimal, and oracle schemes} Figure~\ref{fig:gain_transprob}(a) considers a fixed amount of  total  cache storage $\TC$. The x-axis shows the percentage $\MC/\TC$ of the total cache storage that is allocated to the mid-level cache; hence, 0\% indicates that all cache storage is equally distributed to the leaf caches, whereas 100\% indicates that all cache storage is allocated to the mid-level cache.
Figure~\ref{fig:gain_transprob}(a) shows that  EPC achieves a higher gain than the naive scheme for small values of $\MC/\TC$. Moreover, EPC's gain is close to the optimal scheme, in the case of a flat cache structure ($\MC/\TC=0$). On the other hand, for larger values of  $\MC/\TC$ the gains of all schemes are close, and become equal when all storage is allocated to the mid-level cache; this occurs because the capacity of the mid-level cache is larger than the total demand and all requested objects can be cached, thus the gains for all schemes are equal and determined by the delay  of the mid-level cache.
Figure~\ref{fig:gain_transprob}(a) also shows that the oracle achieves the best performance when all the  storage is allocated to leaf caches, whereas the best performance of the EPC and naive schemes for the specific system parameters is achieved when $\MC/\TC=75\%$.

\mynotex{
\begin{itemize}
\item for different values of MB/TC
\item mention differences in percentage
\item mention performance of eql close to naive.
\item naive: independent of trans probs
\item oracle: also independent of trans probs
\end{itemize}
}

\subsubsection{Influence of mobile transition probabilities}
\begin{figure}[tb]
\centering
\begin{minipage}[]{1\linewidth}
\centering
\includegraphics[width=2.85in] {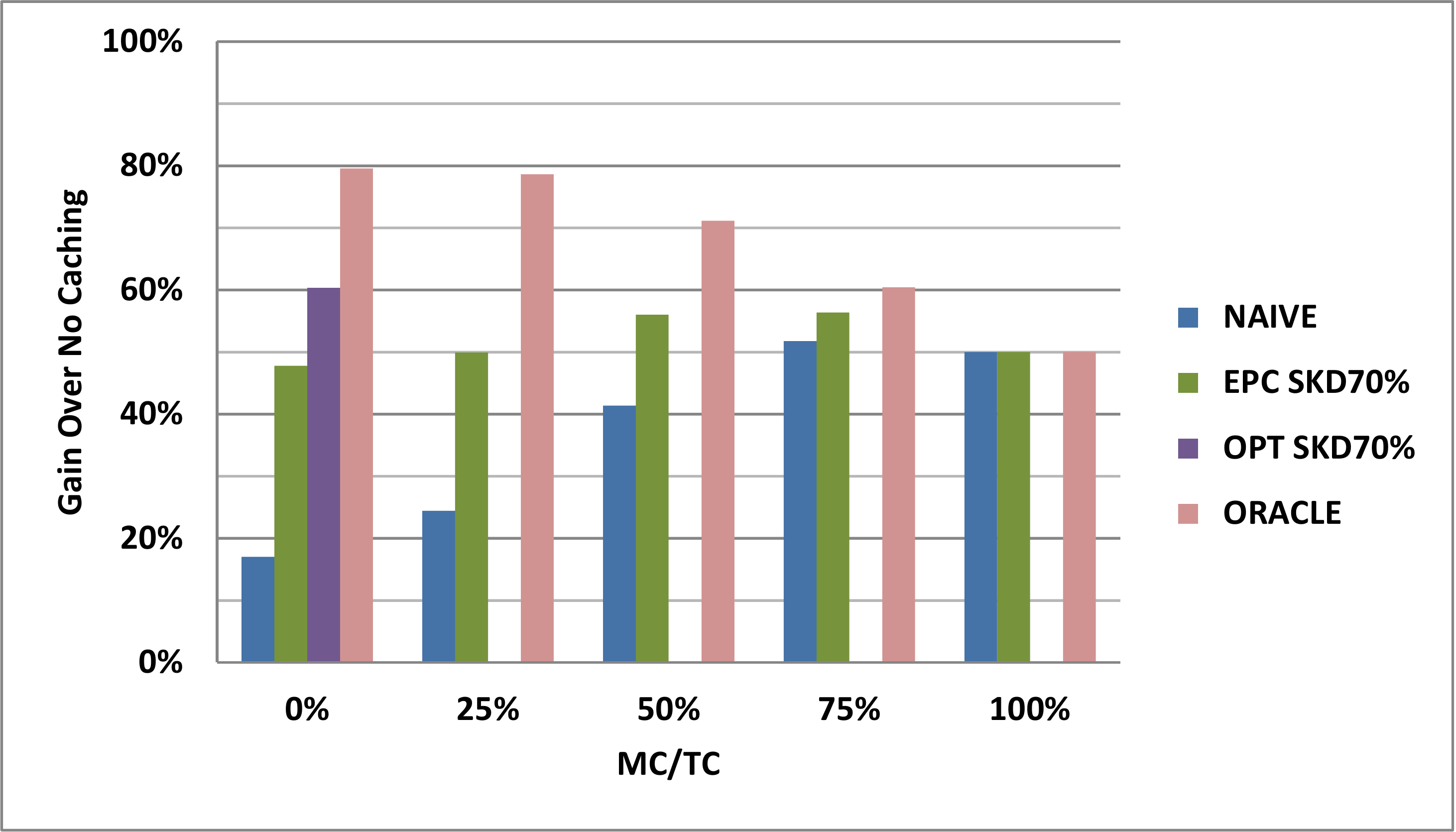} \\
\footnotesize{(a) SKD70\%}
\end{minipage}
\vspace{0.04in}

\begin{minipage}[b]{1\linewidth}
\centering
\includegraphics[width=2.85in] {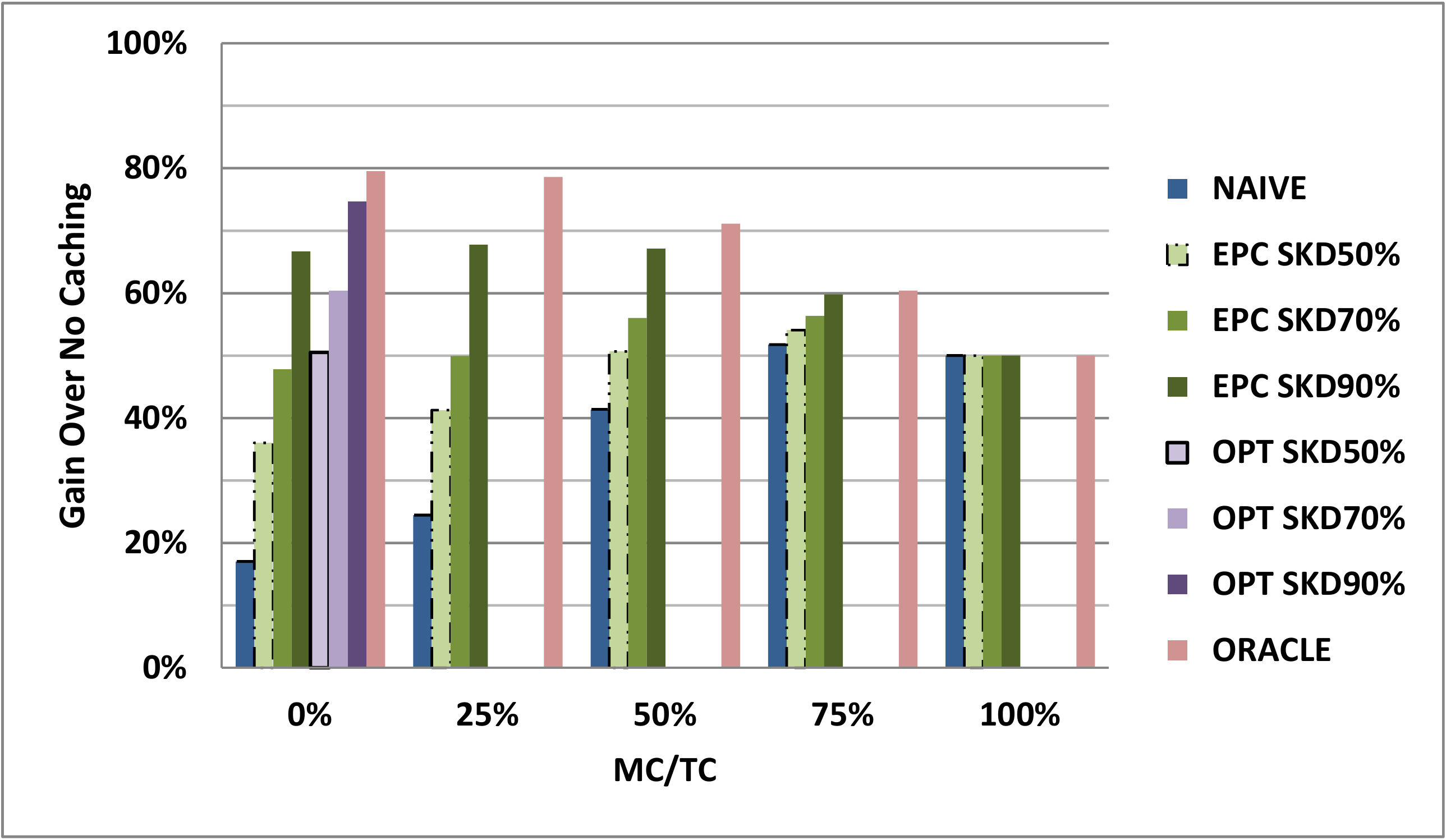} \\
\footnotesize{(b) SKD50\%,70\%,90\%}
\end{minipage}
\vspace{-.2 in}
\caption[]{\protect {Influence of mobile transition probabilities on gain. Fixed-delay, $D_{\R}/D_{\Leaf}=10, D_{\M}/D_{\Leaf}=5$, $\TC=240$ (total cache).}}
\label{fig:gain_transprob}
\vspace{-0.15in}
\end{figure}

Figure~\ref{fig:gain_transprob}(b) shows that when the \emph{skewness} of the mobile transition probabilities increases, then the gains of  EPC are higher when more storage is allocated to the leaf caches.
Moreover,  when the skewness is large (i.e. SKD90\%), then  EPC achieves more than 80\% of the gains achieved by the oracle and almost 90\% of the gains achieved by the optimal scheme for a flat cache structure.
It is important to note that, for large skewness, the EPC scheme achieves  a higher gain than the naive scheme even when the allocation of storage to leaf and mid-caches is selected \emph{to achieve the best performance for each scheme:} Specifically, for $\MC/\TC=25\%$  EPC  achieves  gain 68\%, which is more than 30\% higher than the highest gain achieved by the naive scheme, namely 52\% for $\MC/\TC=75\%$.

\mynotex{
\begin{itemize}
\item more skewed trans probs: improved performance when storage at leaf caches higher.
\end{itemize}
}

\subsubsection{Influence of delay ratios} 
\begin{figure}[tb]
\centering
\begin{minipage}[b]{1\linewidth}
\centering
\includegraphics[width=2.85in] {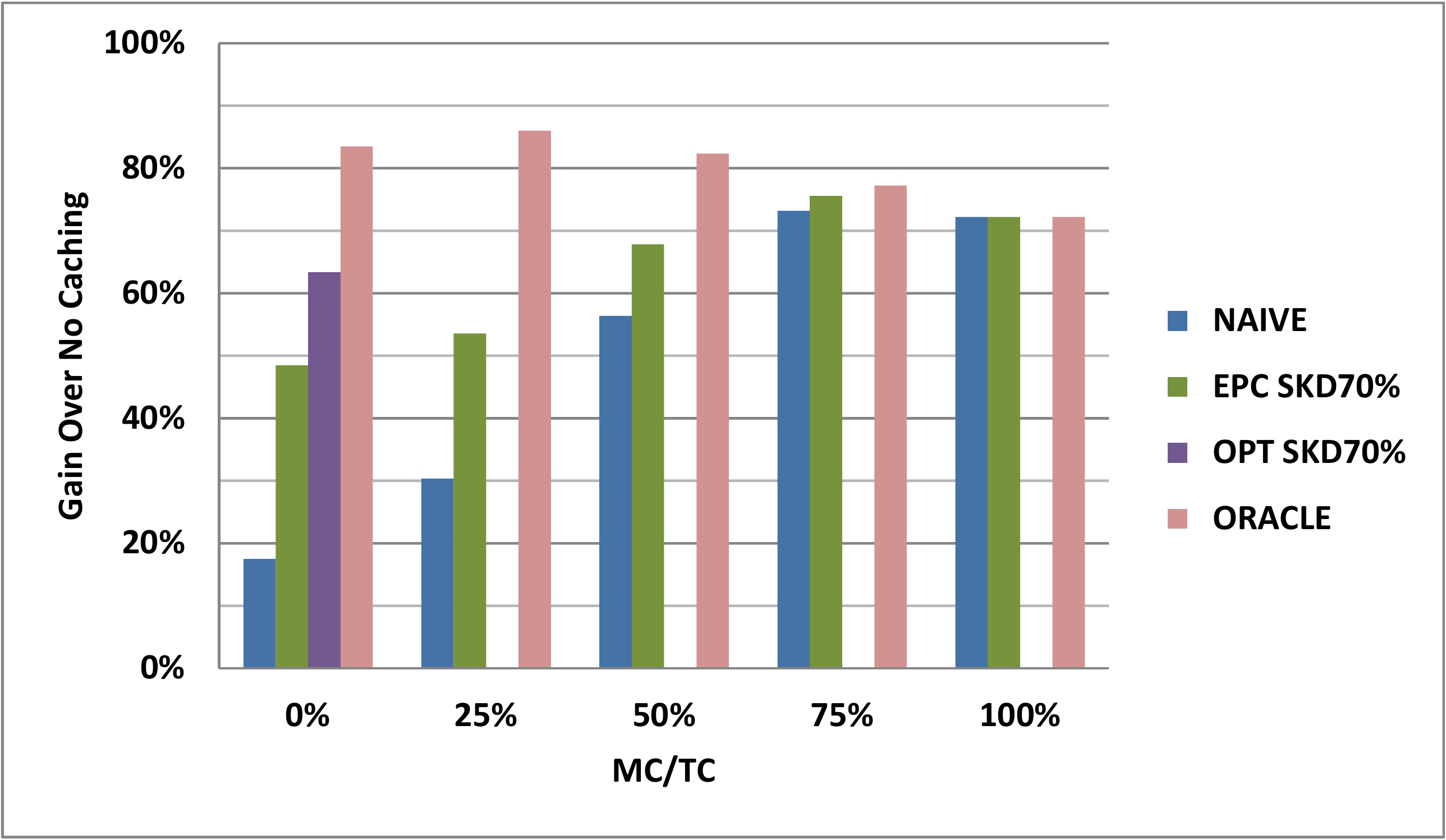}\\
{\footnotesize {(a) $D_{\R}/D_{\Leaf}=18, D_{\M}/D_{\Leaf}=5$}}
\end{minipage}
\vspace{-0.005in}

\begin{minipage}[b]{1\linewidth}
\centering
\includegraphics[width=2.85in]{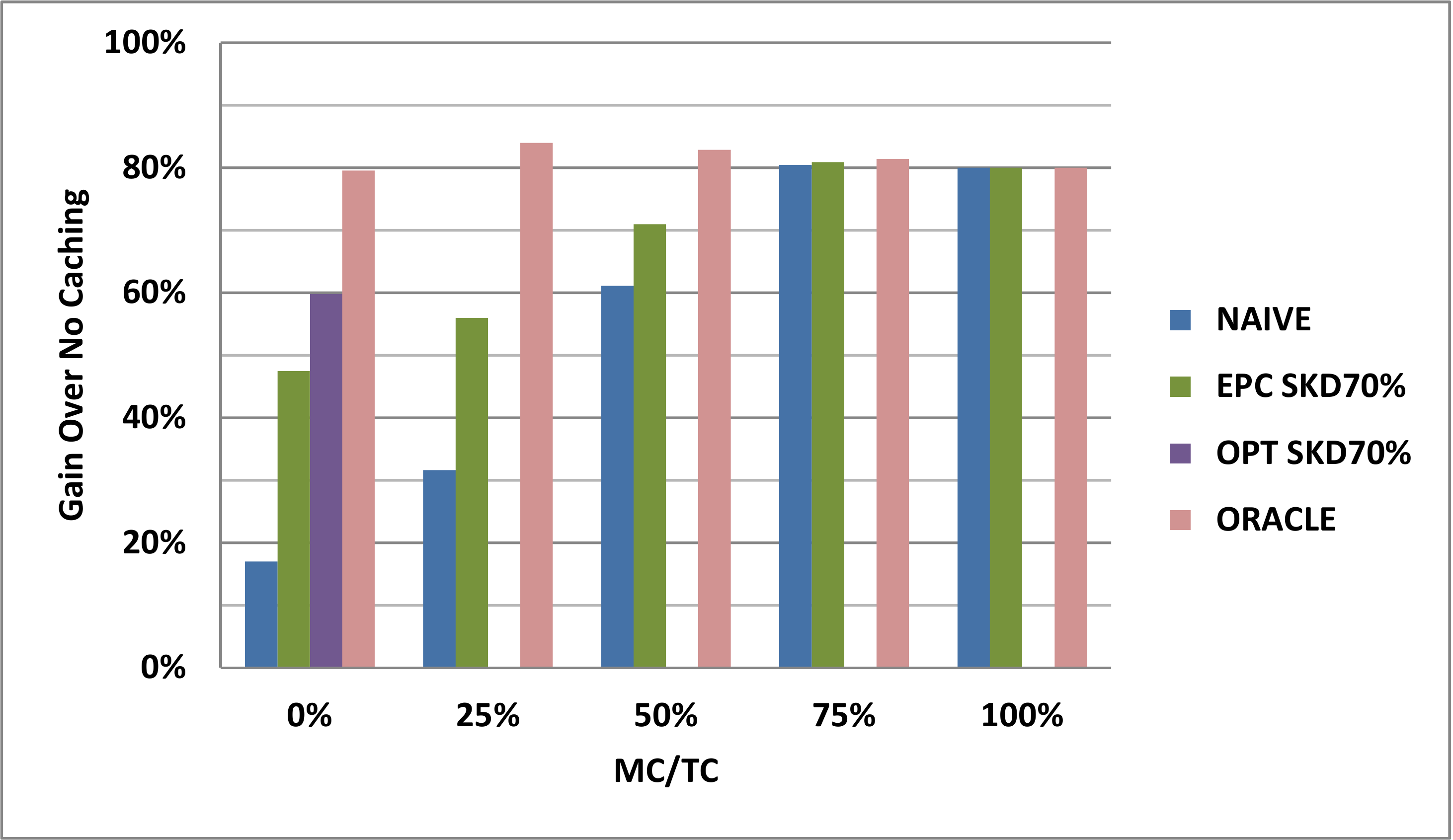}\\
{\footnotesize  {(b) $D_{\R}/D_{\Leaf}=10, D_{\M}/D_{\Leaf}=2$}}
\end{minipage}
\vspace{-.2 in}
\caption[]{\protect {Influence of delay ratios on gain. Fixed delay, SKD70\%, $\TC=240$.}}
\label{fig:gain_delay}
\vspace{-0.2in}
\end{figure}

Comparing Figure~\ref{fig:gain_delay}(a) to \ref{fig:gain_transprob}(a) shows that when the delay for obtaining data objects from their original sources is higher, then the performance of all schemes is generally higher, especially when more storage is allocated to the mid-level cache (larger values of  $\MC/\TC$). Figure~\ref{fig:gain_delay}(b) shows that when the delay for obtaining data from a mid-level cache is close to the delay for obtaining the data from a leaf cache ($D_{\M}/D_{\Leaf}=2$), then EPC achieves the highest  gains when all storage is allocated to the mid-level cache.

\mynotex{
\begin{itemize}
\item 18,5: higher gains when delay from source is higher
\item 10,2: optimal MB/TC approaches 100\%
\end{itemize}
}

\subsubsection{Influence of total cache storage} 
\begin{figure}[tb]
\centering
\begin{minipage}[b]{1\linewidth}
\centering
\includegraphics[width=2.8in] {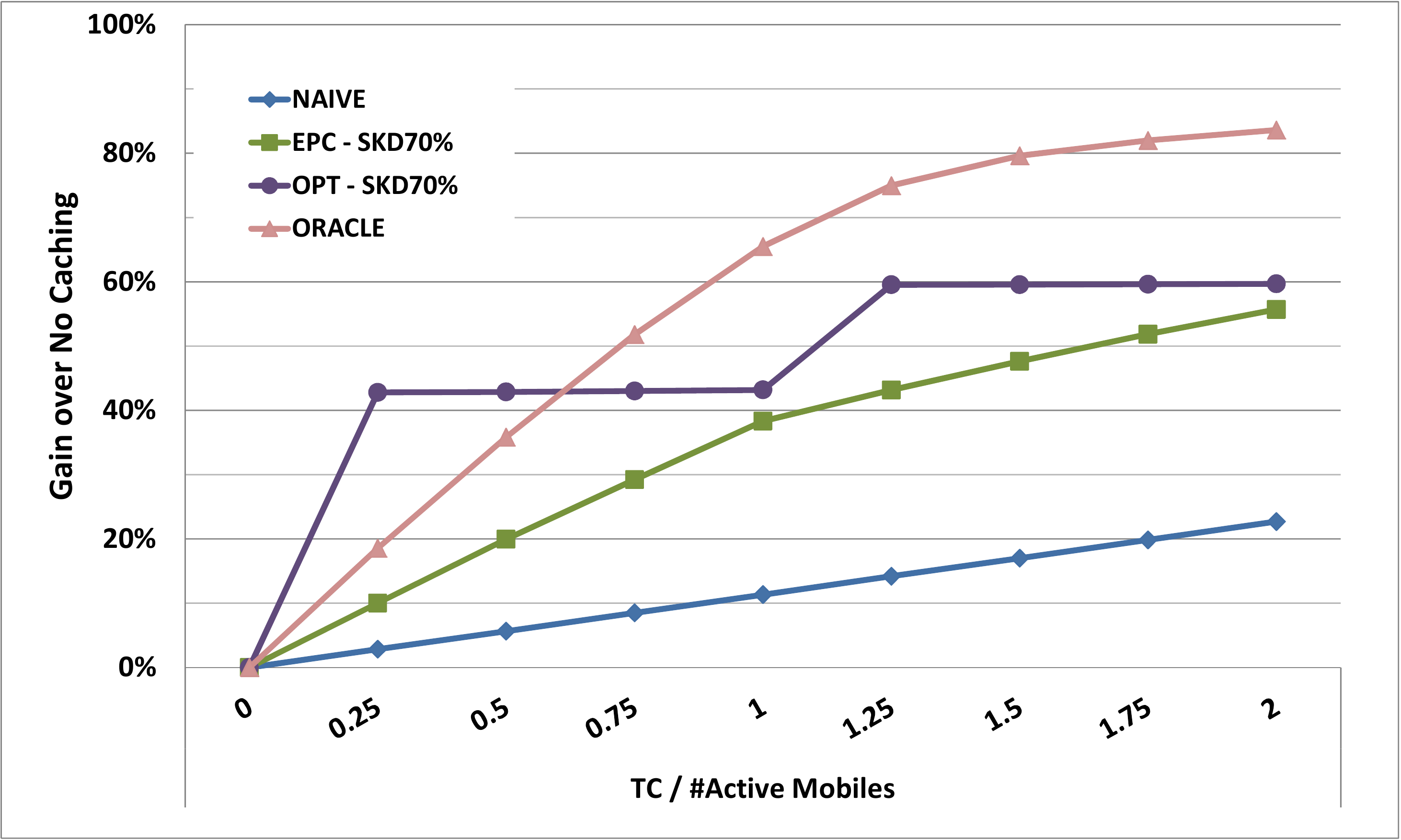}\\
{\footnotesize {(a) $MC=0$ (flat cache structure)}}
\end{minipage}
\vspace{-0.04in}

\begin{minipage}[b]{1\linewidth}
\centering
\includegraphics[width=2.8in]{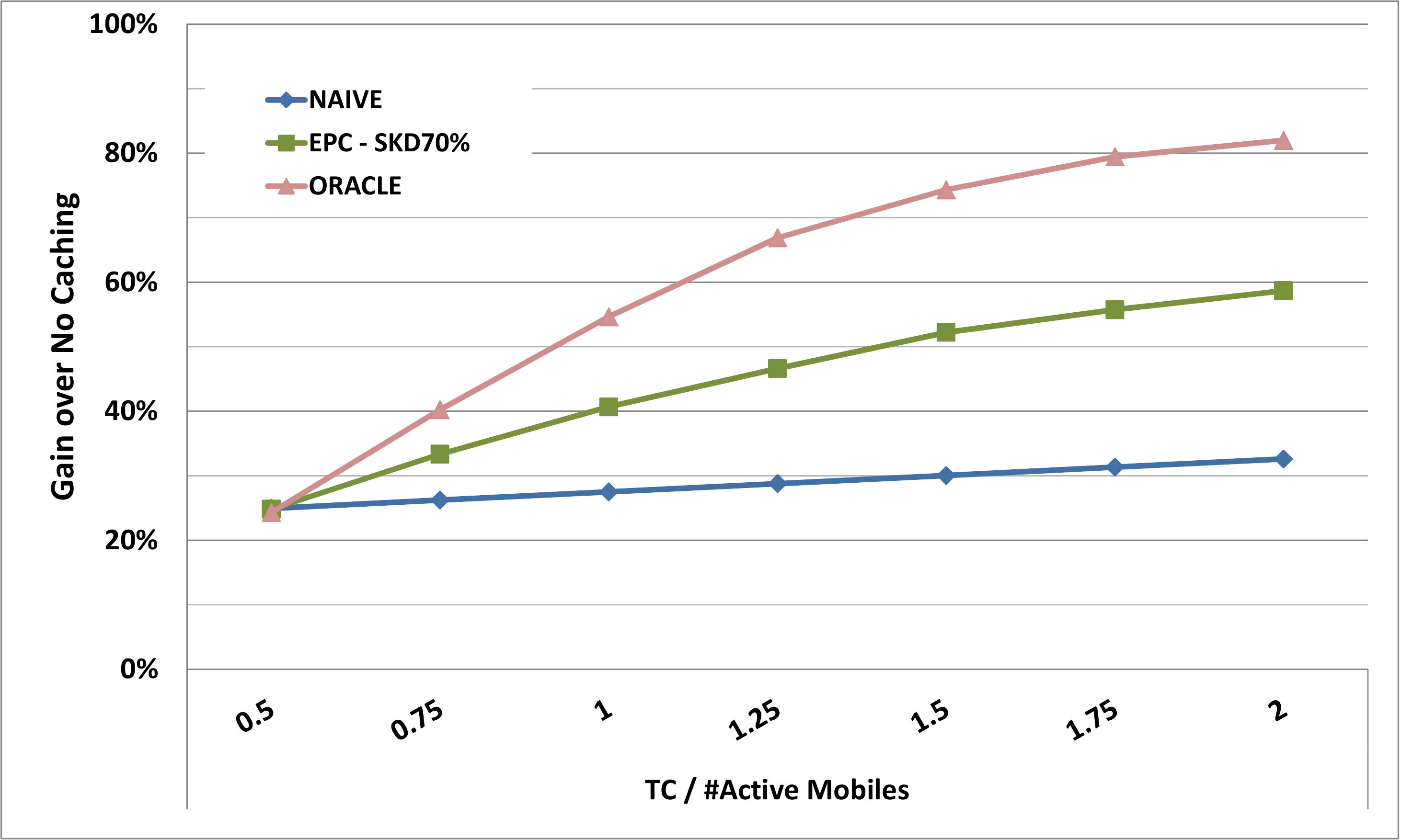}\\
{\footnotesize  {(b) $MC=80$ (mid-level cache)}}
\end{minipage}
\vspace{-.2 in}
\caption[]{\protect {Influence of total cache size on gain. Scaled-down Internet topology.}}
\label{fig:gain_tc}
\vspace{-0.2in}
\end{figure}

Next we consider a scenario with the scaled-down Internet topology. Figure~\ref{fig:gain_tc}(a) shows the influence of the total cache storage on the performance in the case of a flat cache structure, while Figure~\ref{fig:gain_tc}(b) shows the same influence when the mid-level cache of each neighborhood can store 80~objects.
Observe that while the gain of the naive scheme increases linearly with the total cache, the gains for the EPC and oracle increase slower for larger values of the total cache. Furthermore, the optimal scheme exhibits a stepwise behavior, which is a result of its operation: the optimal scheme recomputes the cache allocation  whenever cache storage is freed; this is also the reason that the optimal scheme has, for small values of cache storage, a higher gain than  oracle:  the oracle scheme decides where an object is cached when the corresponding request appears; this decision does not change until the corresponding handoff is performed. However, it may happen that when the decision was made there was no available  storage  at the cache where the mobile will eventually move to. On the other hand, the optimal scheme continuously recomputes the cache allocation when storage space becomes available, hence can take advantage of any free cache space right before a handover is performed.

\mynotex{
\begin{itemize}
\item possibly do this for variable delay
\item why is optimal better than oracle: oracle caches an object as soon as it appears and before the handover. Optimal: recomputes optimal each time cache storage freed, considering all the active cache requests.
\item concave shape.
\end{itemize}
}

\subsubsection{Transient behavior} 
\begin{figure}[b]
\vspace{-0.2in}
\centering
\includegraphics[width=2.8in]{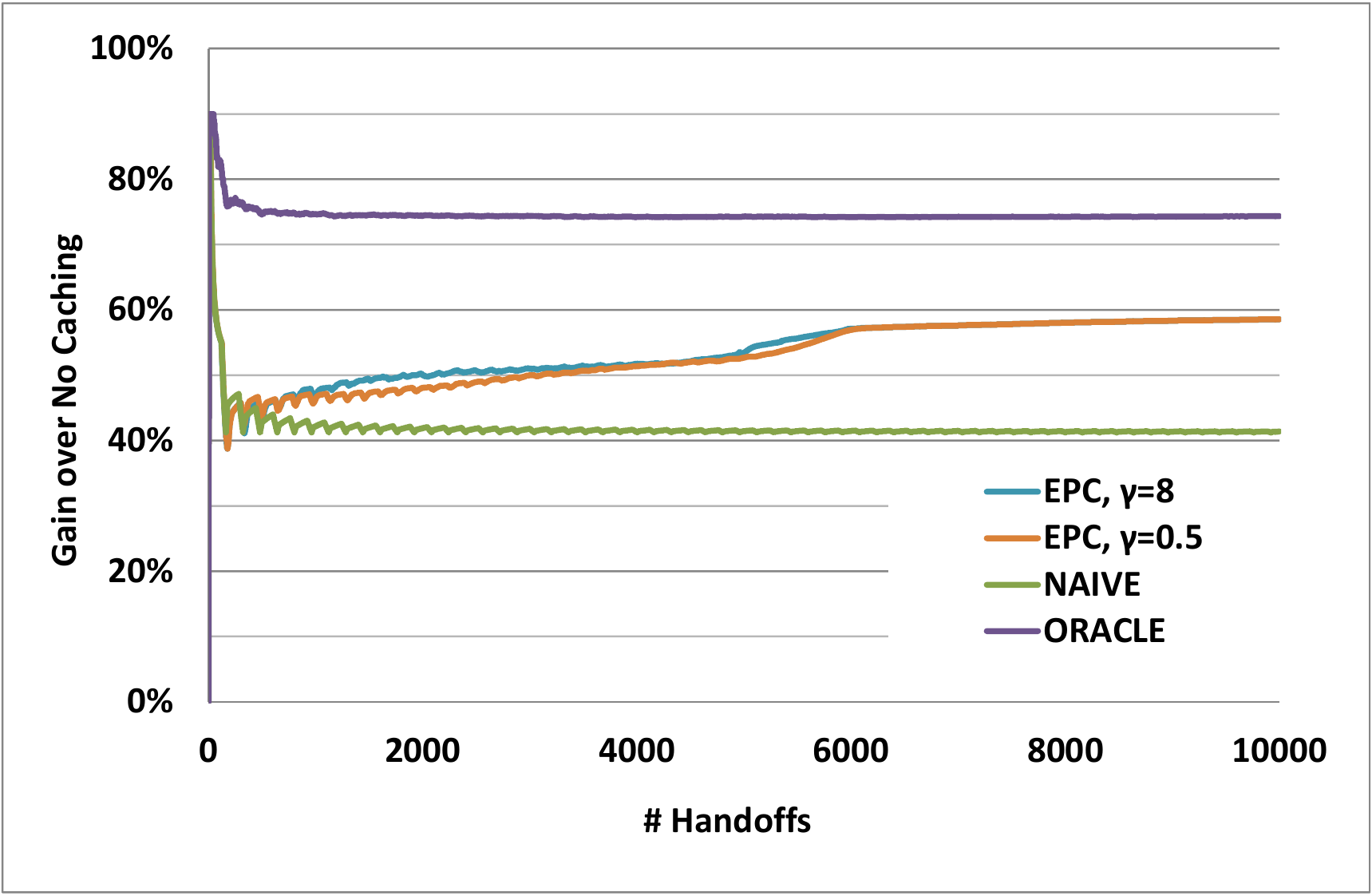}
\vspace{-0.08in}
\caption{Transient behavior. The EPC scheme converges slower for smaller values of the price update factor $\gamma$. At point 5000 the mobile transition probabilities change from SKD70\% to SKD90\%. Fixed delay.}
\label{fig:trans}
\end{figure}

Figure~\ref{fig:trans} shows that when the system starts with no knowledge of the mobile transition probabilities, EPC's gain converges to its steady state value slower than the other schemes; this is due to the dynamic updating of  cache congestion prices and the measurement-based estimation of  mobile transition probabilities. Moreover, as expected, the convergence time is larger for smaller values of  factor $\gamma$: after a change of the mobile transition probabilities from SKD70\% to SKD90\% at time 5000, the gain reaches 95\% of its steady state value after approximately 750 handovers for $\gamma=0.5$ and 490 handovers for $\gamma=8$.

\mynotex{
\begin{itemize}
\item how is gain estimated in the transient graph: average of delays then take ratio to get gain
\item identify when 95\% of steady state value is reached: approx 750 for $\gamma=0.5$ and 490 for $\gamma=8$
\end{itemize}
}

\subsubsection{Influence of variation of mobile transition probabilities} 
\begin{figure}[tb]
\centering
\includegraphics[width=2.85in]{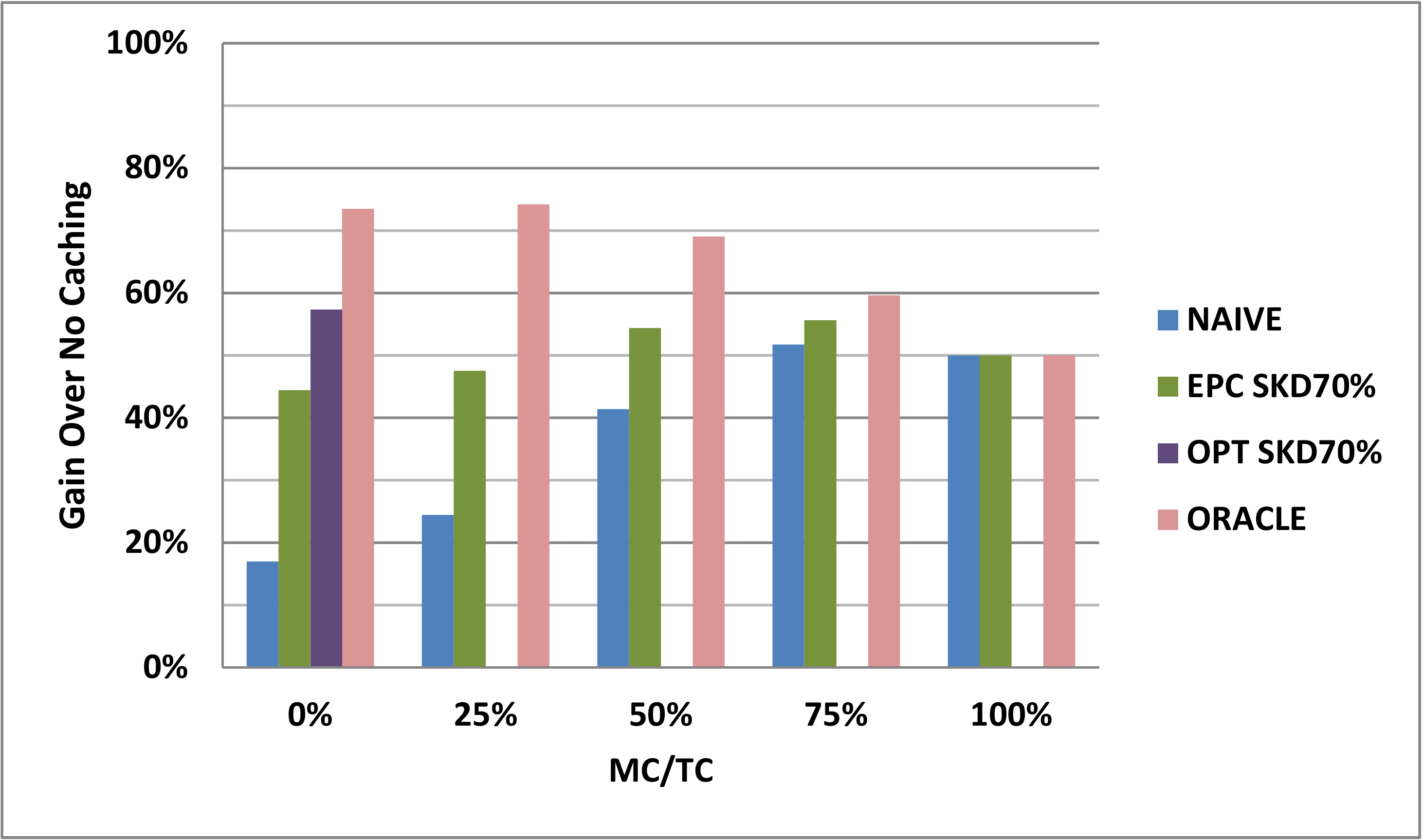}
\vspace{-0.1in}
\caption{Influence of variation of mobile transition probabilities. Fixed delay, standard deviation of mobile transition probabilities $30\%$.}
\label{fig:gain_var_trans}
\end{figure}

Comparison of Figures~\ref{fig:gain_var_trans} and \ref{fig:gain_transprob}(a) shows that  EPC's   gains  are robust to the variation of the mobile transition probabilities: for values of $MC$ that EPC achieves the highest gains (50\% and 75\%), an increase of the standard deviation of the  transition probabilities from 5\% (Figure~\ref{fig:gain_transprob}(a)) to 30\% (Figure~\ref{fig:gain_var_trans}) reduces the gains of EPC by less than 10\%.
 Moreover, the higher variation of the  transition probabilities appears to influence the oracle  more than the other schemes.
This is because the higher variation can lead the oracle to require more leaf-level caching than what is  available, hence is forced to use the mid-level cache which achieves lower delay gains; this is verified by the higher use of the mid-level buffer by the oracle, for a higher variation of the transition probabilities.

\mynote{
\begin{itemize}
\item Add note that this is verified by the increased utilization of the mid-level cache for the oracle.
\end{itemize}
}

\begin{figure}[tb]
\vspace{-0.05in}
\centering
\includegraphics[width=2.85in]{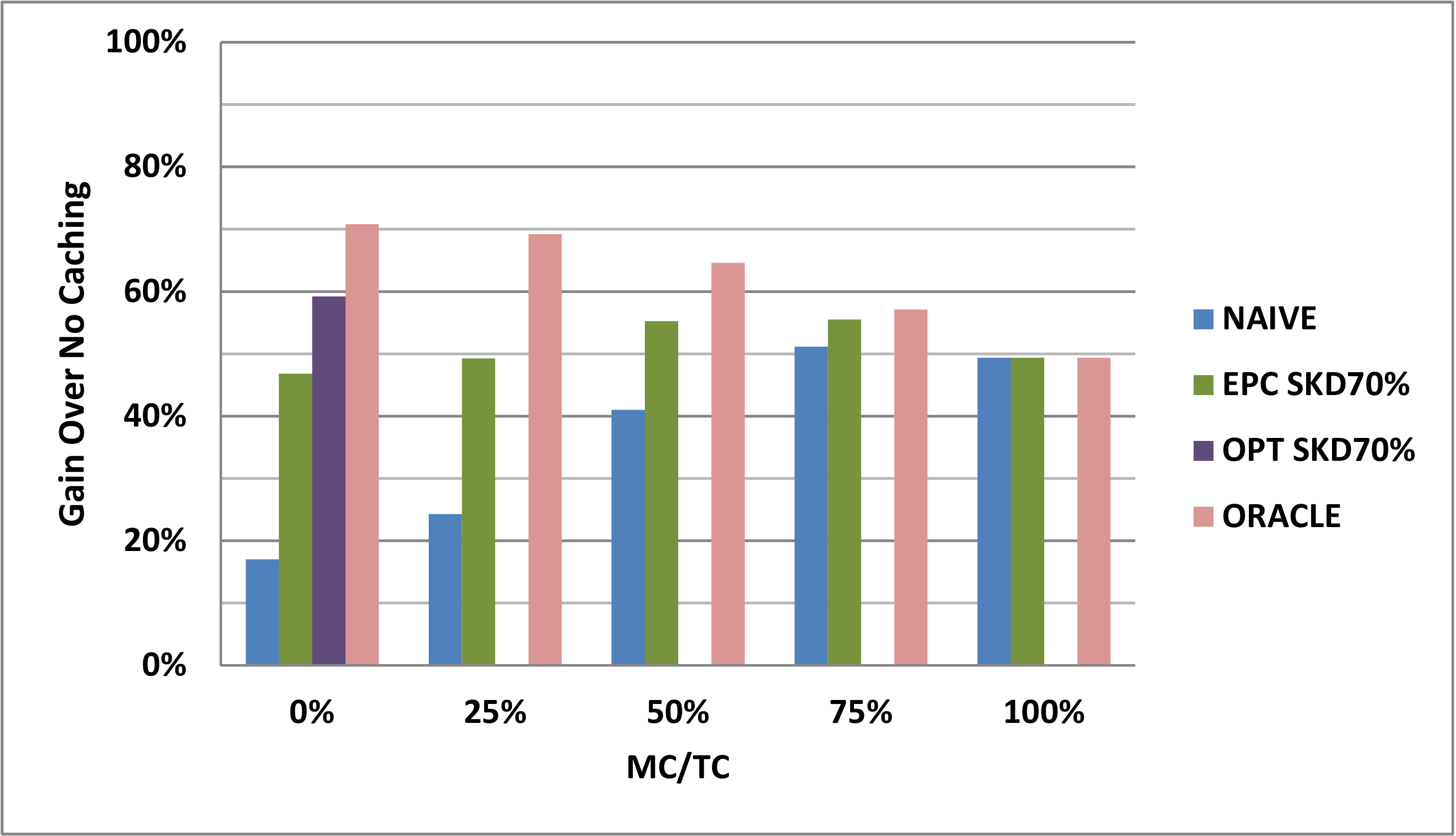}
\vspace{-0.1in}
\caption{Scaled-down Internet topology, $D_{\R}/D_{\Leaf}=10, D_{\M}/D_{\Leaf}=5$, $\TC=240$ (total cache).}
\label{fig:gain_70_int}
\vspace{-0.3in}
\end{figure}

\subsubsection{Comparison of fixed  delay with scaled-down  Internet topology} Comparison of Figure~\ref{fig:gain_70_int}, which is for the scaled-down Internet topology, and  Figure~\ref{fig:gain_transprob}(a), which is for fixed delay, shows that the variation due to the Internet topology has a larger influence on the oracle scheme; similar to the above explanation for the higher variation of the mobile transition probabilities, this is because the higher variation results in the oracle scheme  requiring leaf-level caching more than the available storage, thus forcing it to use the mid-level cache for which the delay gains are lower.

\mynote{
\begin{itemize}
\item need to check/verify this explanation.
\end{itemize}
}

\vspace{-0.09in}
\section{Conclusions}
\label{sec:conclusions}
\vspace{-0.03in}

We have presented and evaluated a proactive caching scheme for reducing the delay in mobile scenarios, which exploits mobility information and uses congestion pricing  to efficiently utilize cache storage. Our modeling framework includes the case of a  flat cache structure and a two-level cache hierarchy, and  the case where the delay is independent of the size of the requested objects and  where the delay is a function of the object sizes. Our evaluation results
show how various parameters influence the delay gains of the proposed  scheme, which achieves robust and good performance relative to a scheme which attempts to naively cache all data objects,  an optimal scheme for a flat cache structure, and an oracle which knows a priori a mobile's future  attachment point.

Ongoing and future work includes extending the proactive caching scheme to consider cases where the same data object is requested by more than one mobile user, thus exploiting mobility information together with  object popularity.
A second extension  is to allow a mobile, after a cache miss, to obtain the requested data object not from the original source, but from another cache that proactively fetched the object and is closer to the mobile than the original source; this corresponds to a hierarchy with two or more levels of caches, which can be further motivated when inter-ISP charging costs are present.
The proposed models can also be extended to include constraints on the available capacity, such as  a low capacity backhaul in femto/small cells  and WiFi hotspots, thus capturing in a uniform manner the constraints on cache storage and network capacity.
Another direction for future work involves adapting the proposed proactive caching framework that uses congestion pricing to achieve efficient cache utilization to  dense femto/small cells and WiFi hotspots with overlapping coverage. The content placement problem in such topologies is NP-hard \cite{Gol++12}, hence a solution like the one proposed in this paper that uses congestion pricing can have advantages.
Finally, an interesting direction is the application of proactive caching and congested resource pricing in mobile software agent scenarios, where mobility involves software relocation rather than device/physical mobility. In this context, the constrained resources can be storage, for delay critical applications that process a large amount of data, or CPU processing.

\mynote{
\begin{itemize}
\item Although problem is different from replication and traditional caching, a congestion pricing approach to reflect cost of caching can also be applied. In particular, we can extend the models in the paper by assuming  that the same data object is requested by more than one mobiles.
\item take into account popularity, hence case where mobiles ask for same objects.
\item application of efficient proactive caching approach to case of dense WLANs with overlapping coverage, \cite{Gol++12}.
\item Utility model can be extended to account for link (WiFi, backhaul) congestion. This may not be that exciting. On the other hand, a motivation can be that we uniformly address both storage and capacity constraints.
\end{itemize}
}

\bibliographystyle{IEEEtran}

\vspace{-0.09in}
\bibliography{proactive_caching}

\end{document}